\documentclass[aps,prd,showpacs,twocolumn,superscriptaddress]{revtex4}

\usepackage{graphicx}% Include figure files
\usepackage{dcolumn}% Align table columns on decimal point
\usepackage{bm}% bold math
\usepackage{color}
\usepackage[normalem]{ulem} % \sout{old text} for strikeout
\usepackage[dvipsnames]{xcolor} % For blue in-text comments 
\usepackage{hyperref}
\hypersetup{
  colorlinks=true,        % false: boxed links; true: colored links
  linkcolor=blue,         % color of internal links
  citecolor=blue,         % color of links to bibliography
}
\usepackage{doi}
\usepackage{amsmath}
\usepackage{amssymb}

\begin{document}
%\preprint{ASTRO-03/2016}

\title{Pulsar magnetospheres in dynamical Chern–Simons gravity: \\deathline conditions and polar-cap particle acceleration} 

\author{Sherzod Sayfiyev}
\email{sayfiyev-sherzod@samdu.uz}
\affiliation{School of Physics, Harbin Institute of Technology, Harbin 150001, China}
\affiliation{Samarkand State University, University Blvd.15, Samarkand 140104, Uzbekistan}
\affiliation{Tashkent State Technical University, Tashkent 100095, Uzbekistan}

\author{Bobomurat~Ahmedov}
\email{corresponding author: ahmedov@astrin.uz}
\affiliation{School of Physics, Harbin Institute of Technology, Harbin 150001, China}
\affiliation{Institute of Theoretical Physics, National University of Uzbekistan, Tashkent 100174, Uzbekistan}
\affiliation{Institute for Advanced Studies, New Uzbekistan University, Movarounnahr str. 1, Tashkent 100000, Uzbekistan}

\author{Chengxun Yuan}
\email{yuancx@hit.edu.cn}
\affiliation{School of Physics, Harbin Institute of Technology, Harbin 150001, China}

\author{Javlon Rayimbaev}
\email{javlon@astrin.uz}
\affiliation{University of Tashkent for Applied Sciences, Gavhar Str. 1, Tashkent 700127, Uzbekistan}
\affiliation{Kimyo International University in Tashkent, Shota Rustaveli Street 156, Tashkent 100121, Uzbekistan}

\author{Bahodir~Ahmedov}
\email{b.ahmedov@newuu.uz}
\affiliation{New Uzbekistan University, Movarounnahr Street 1,  Tashkent 100000, Uzbekistan}

\author{Asadbek Shermanov} \email{shermanovasadbek@gmail.com} \affiliation{Institute of Theoretical Physics, National University of Uzbekistan, Tashkent 100174, Uzbekistan}

\date{\today}

\begin{abstract}
The role of surface gravity in neutron stars (NSs) is considerable in the radiation mechanisms of the surrounding plasma magnetosphere near the star surface, as they are highly magnetized, compact gravitating objects with compactness $M/R \simeq 0.2$. 
In this context, they serve as a celestial laboratory for testing gravity theories that dominate the stars' exteriors through plasma magnetospheric radiation. In this paper, we examine how dynamical Chern-Simons (dCS) gravity modifies pulsar electrodynamics in the slow-rotation, weak-coupling regime; the leading-order correction affects only the off-diagonal metric component $g_{t\phi}$, while the diagonal components remain as in General Relativity (GR). We first solve the Maxwell equations in the dCS framework for the electromagnetic field components and obtain analytical solutions for the electric and magnetic fields. We show that the magnetic field components are the same as in GR, while the electric field components are modified in dCS. We then derive analytic expressions for the induced electric charge density, known as the Goldreich–Julian (GJ) charge density, which sources the induced electric field arising from the magnetic field and the star's rotation. In dCS, it is larger than in the GR case, whereas the accelerating electric field parallel to the magnetic field lines is weaker. Next, we consider the deathline condition for switching off electron and positron radiation via inverse Compton scattering in the star's polar cap zone and show that the line shifts upward in the $P-\dot{P}$ diagram, explaining the physics of shorter-lived pulsars.
Lastly, we study charged-particle acceleration in the polar cap region and show that electrons reach higher energies over shorter distances than in the GR case.
\end{abstract}


\maketitle

\section{Introduction}

The electrodynamics of rotating neutron stars (NS) remains one of the most direct ways to probe the role of NS gravity on plasma magnetospheric radiation processes. In the general-relativistic (GR) framework \cite{Muslimov1992, Turimov2021}, the structure of the pulsar magnetosphere is governed by frame dragging, which determines the charge density proposed by Goldreich \& Julian (GJ) \cite{Goldreich1969} known as the GJ charge density, the accelerating electric field, and ultimately the conditions for particle acceleration and pair creation \cite{Harding1998, Harding2002}. The classical formulation developed by Muslimov and Tsygan established that even small relativistic corrections to frame dragging lead to significant changes in the electrodynamic structure of the polar-cap region \cite{Muslimov1992}. Although NSs were predicted soon after the discovery of neutrons in 1932, intensive studies of NS physics received considerable attention only after their discovery as radio pulsars in astronomical observations in 1967~\citep{Hewish68}. The most important issue at the time was determining the source of pulsars' electromagnetic radiation. For the first time, it was proposed that the electromagnetic radiation from radio pulsars causes loss of rotational kinetic energy \citep{Pacini68,Gold68}.

Studies of compact objects in Chern-Simons (CS) gravity have established black holes and NSs as clean probes of parity-violating corrections to GR. Refs.~\cite {JackiwPi2003, AlexanderYunes2009} clarified the formulation of CS gravity and its dynamical extension, while Grumiller and Yunes showed that the usual stationary Kerr ansatz is strongly restricted in the nondynamical theory \cite{GrumillerYunes2008}. Approximate rotating black-hole solutions were then obtained in the slow-rotation/small-coupling regime, showing that the dominant correction enters the gravitomagnetic sector and modifies frame dragging \cite{YunesPretorius2009,KonnoMatsuyamaTanda2009}. Perturbative studies of Schwarzschild and slowly rotating black holes demonstrated axial--scalar mode coupling, modified quasinormal spectra, and stability within the effective regime \cite{CardosoGualtieri2009,MolinaPaniCardosoGualtieri2010,YagiYunesTanaka2012}. These solutions were further extended toward rapid rotation, extremality, nonperturbative numerical spacetimes, and binary black-hole collisions, revealing coupling-dependent deviations in the scalar field, quadrupole structure, waveforms, and remnant dynamics \cite{Stein2014,McNeesSteinYunes2016,DelsateHerdeiroRadu2018,OkounkovaSteinScheelTeukolsky2019}. For NSs, CS corrections were shown to leave the leading mass--radius relation essentially unchanged but to modify the gravitomagnetic field and moment of inertia \cite{YunesPsaltisOzelLoeb2010,AliHaimoudChen2011}. Later work constructed isolated and binary neutron-star solutions and universal I--Love--Q relations, indicating that future electromagnetic and gravitational-wave measurements may constrain the CS coupling far more strongly than weak-field tests \cite{YagiSteinYunesTanaka2013,YagiYunes2013,GuptaMajumderYagiYunes2018}.

The first study of electromagnetic fields of a dipolar magnetized sphere in vacuum (near and wave zones) is presented in Ref.\cite{Deutsch1955}. Whereas gravity at the surfaces of NSs is considerable, they are relativistic laboratories for testing gravity. These significant features led researchers to study the effects of Einstein's GR and alternative/modified gravity theories on the electrodynamics of magnetized NSs ~\cite{Muslimov1986SvA,Konno00,Rezzolla01c,Rezzolla01d,Fattoyev2008PhRvD,Hakimov13,Turimov17,Turimov18a,Rayimbaev2020b}. Observational signals from pulsars that arise from radiation processes in the NS plasma magnetosphere along open field lines explain the electromagnetic radiation of NSs from radiative processes involving accelerated particles \citep{Sakai03,Beloborodov08,Rayimbaev2019IJMPD,Morozova08,Rayimbaev2020b}. According to the rotation-powered NS model \cite{Pacini68,Gold68}, the star slows down, in which the kinetic energy of star rotation is converted into electromagnetic radiation of radio pulsars, and the observed spin-down rate of radio pulsars, which is measured with high precision, is related to kinetic energy losses ~\cite{Cheng1986a,Cheng1986b}. The decay of the stellar magnetic field due to the high conductivity of the stellar matter can be negligible. The decrease in angular velocity reduces the accelerating parallel electric field near the surface, thereby breaking the chain reaction of pair production \cite{Daugherty1982,Daugherty1996} in plasma magnetosphere formation. Thus, the pulsar can no longer radiate and disappears from radio observations \cite{Ruderman75}. This phenomenon is known as the radio pulsar death line in $B-P$ or $P-\dot{P}$ spaces ~\cite{Chen93,Kantor04,Morozova12,Zhang96a,Zhang2004,AhmedovMorozova2012,Rayimbaev2019,Rayimbaev2019IJMPD,Rayimbaev2020,Sayfiyev2025,Rayimbaev2020b,Bokhari2021PDU}

The central aim of the present work is to show how dCS modifies pulsar electrodynamics. Therefore, we focus on dCS gravity, where, in the slow-rotation, weak-coupling regime, the leading-order correction modifies only the off-diagonal metric component $g_{t\phi}$, while the diagonal components remain identical to those of GR. From this point, dCS gravity is fundamentally different from many alternative/modified gravity models. Here, the magnetic-field geometry is preserved, and all deviations arise purely from the modified frame-dragging effect. %We demonstrate that a decrease in the frame-dragging frequency increases the effective angular velocity difference, thereby enhancing the deviation of space and GJ charge density in the near-surface region. In turn, the accelerating electric field is reduced due to the global structure of the electrostatic potential, showing that the charge imbalance and the accelerating field are controlled by different aspects of the system. This separation of effects represents one of the key physical insights of the present work.

%The main novelty of this work is that we solve, in a self-consistent analytical form, the electromagnetic field structure and the plasma-filled magnetosphere of a slowly rotating neutron star in dCS gravity, including the GJ charge density, the electrostatic potential, the accelerating electric field, and the pulsar death line. By keeping the magnetic-field configuration unchanged and modifying only the frame-dragging term, we can clearly identify the physical origin of all deviations from the GR case.

%A key point of the analysis is the use of a slowly rotating black-hole solution in dCS gravity as an effective exterior spacetime for a neutron star. At first sight, this may appear restrictive, since the exact neutron-star solution in dCS gravity depends on the interior structure and matching conditions at the stellar surface. However, the present approach is justified under well-defined assumptions. First, we work in the slow-rotation approximation, where the spacetime can be treated perturbatively. Second, we restrict ourselves to the exterior region $r \geq R$, where the magnetospheric processes take place. Third, and most importantly, the leading-order dCS correction enters only through the gravitomagnetic sector, which is determined by the star's global angular momentum and is largely insensitive to the star's detailed interior structure at this order. Therefore, the adopted metric provides a consistent and physically well-motivated approximation for studying polar-cap electrodynamics in the exterior spacetime.

The paper is organized as follows. In Sec.\ref{Sec2}, we introduce the dCS-corrected spacetime of a slowly rotating NS and describe the electromagnetic field structure in the ZAMO frame. In Sec. \ref{Sec3}, we derive the GJ charge density and analyze the geometry of the polar-cap region. In Sec. \ref{Sec4}, we solve the Poisson equation for the electrostatic potential and obtain analytical expressions for the accelerating electric field in both the near-surface and far-zone limits. In Sec. \ref{Sec5}, we formulate the condition for electron–positron pair production, derive the pulsar death line, analyze its dependence on the dCS parameter, and compare it with observational data. In Sec. \ref{Sec6}, we investigate particle acceleration by solving the equations of motion and examining Lorentz-factor evolution in different regimes. Finally, in Sec. \ref{Summary}, we summarize the main results and discuss their physical implications.

Throughout this paper, we use the metric signature $(-, +, +, +)$ and work in geometrized units with $G = 1$ and $c = 1$. Latin indices ($i, j, \ldots$) run from 1 to 3 and denote spatial coordinates, while Greek indices ($\mu, \nu, \ldots$) run from 0 to 3 and refer to spacetime components.

\section{Vacuum solutions of Maxwell equations in dCS gravity \label{Sec2}}

\subsection{Slowly rotating neutron-star metric in dCS gravity}
We consider the exterior spacetime of a slowly rotating NS in dCS gravity. In the slow-rotation regime ($a \ll 1$ \cite{Turimov2021,Juraeva2022ArabJMath}) and weak coupling, the metric can be treated perturbatively, and the leading-order correction arises solely in the gravitomagnetic sector of spacetime~\cite{YunesPretorius2009,YagiYunes2013}. This property follows from the parity-violating nature of the CS coupling, which affects only the axial part of the gravitational field.
In this approximation, the diagonal components of the metric remain as in GR, while $g_{t\phi}$ acquires an additional correction induced by the coupling between the scalar field and the Pontryagin density. As a result, the spacetime can be represented as a deformation of the slowly rotating GR solution, with modifications entering exclusively through the frame-dragging term~\cite{YunesPretorius2009}. In Boyer--Lindquist-type coordinates \(x^\mu=(t,r,\theta,\phi)\), the line element can be written as
\begin{equation} \label{metric_dcs}
    ds^2 = ds^2_{\rm GR} + \frac{20 \pi\alpha^2 a}{r^4}
    \left(1+\frac{12M}{7r}+\frac{27M^2}{10r^2}\right)\sin^2\theta\, dt\, d\phi ,
\end{equation}
where $ds^2_{\rm GR}$ denotes the slowly rotating spacetime in GR and the second term represents the leading-order dCS correction to the gravitomagnetic component of the metric. In the GR case, the slowly rotating metric is given by
\begin{eqnarray}\label{metric_sr}\nonumber
    ds^2_{\rm GR} &=&
-N^2(r)\,dt^2 + \frac{dr^2}{N^2(r)} + r^2 d\theta^2 + r^2\sin^2\theta\, d\phi^2 \\
&\quad& - 2\omega(r) r^2 \sin^2\theta\, dt\, d\phi \, ,
\end{eqnarray}
with
\begin{equation} \label{lapse_dragging}
    N^2(r)=1-\frac{2M}{r}, \qquad \omega(r)=\frac{2aM}{r^3}, \qquad r\ge R,
\end{equation}
where $M$ and $R$ denote the stellar mass and radius, $J$ is the total angular momentum, and $a=J/M$ is the specific angular momentum, following the standard slow-rotation formalism introduced by Hartle~\cite{Hartle1967}. In this framework, the function $\omega(r)$ represents the angular velocity of local inertial frames and plays a central role in neutron-star electrodynamics.
The additional term in Eq.~(\ref{metric_dcs}) originates from the coupling between the scalar field and the Pontryagin density in the dCS action. As shown in Refs.~\cite{YunesPretorius2009,YagiYunes2013}, this correction modifies only $g_{t\phi}$ at leading order, while leaving the function $N(r)$ as in GR. Consequently, gravitational redshift and radial structure remain identical to those in GR, and the only modification relevant to magnetospheric processes is a change in frame-dragging.
From a physical point of view, this implies that the angular velocity of inertial frames is altered according to $\omega(r) \rightarrow \omega_*(r)$, which directly affects the corotation condition in the magnetosphere,
\begin{eqnarray}
   \omega_*(r)=\omega(r)\left[1-\frac{10 \pi\alpha^2}{r^3M}
    \left(1+\frac{12M}{7r}+\frac{27M^2}{10r^2}\right)\right]\ .
\end{eqnarray}

%Since the GJ charge density and the induced electric field depend on the difference $(\Omega - \omega)$, any modification of $\omega(r)$ leads to a direct change in the electrodynamic structure of the polar-cap region.
In the nonrotating limit ($a\to0$), the Pontryagin density vanishes identically, and the scalar field in CS gravity becomes trivial. As a result, the exterior spacetime reduces to the GR case.
We use the metric (\ref{metric_dcs}) as an effective description of the exterior region $r \ge R$, where magnetospheric processes take place. In dCS gravity, the exact exterior solution for the NS depends on the internal structure and matching conditions at the stellar surface, and therefore does not coincide exactly with the corresponding black-hole solution~\cite{AliHaimoudChen2011}. However, in the slow-rotation approximation, the frame-dragging term depends primarily on the star's global angular momentum and is only weakly sensitive to interior details at leading order.
This approach is consistent with the standard treatment of NS electrodynamics in GR ~\cite{Rezzolla2001,Muslimov1992}, as well as its extensions to modified gravity models~\cite{Ahmedov2012,Turimov17,Rayimbaev2021}, where the exterior spacetime dominates the magnetosphere structure. From this perspective, the adopted spacetime metric provides a justified and analytically tractable framework for deriving the GJ charge density, the accelerating electric field, and the pulsar death line in dCS gravity.

\subsection{Magnetic field solution}

We now describe the exterior magnetic field of a slowly rotating magnetized NS in the locally nonrotating ZAMO frame using the dipolar approach. The ZAMO frame is naturally adapted to stationary and axisymmetric spacetimes. In this frame, the electromagnetic field components correspond to locally measured physical quantities.

In this work, we consider the dCS spacetime exterior to NSs. The vacuum Maxwell equations that determine the exterior dipolar magnetic field retain the same radial structure as in GR. The NS's slow rotation remains, but its leading electric effect enters through frame dragging and the induced electric field, not through a deformation of the dipolar magnetic geometry \cite{YunesPretorius2009, YagiYunes2013}.
This is why the magnetic-field components may be taken from the GR solution for a slowly rotating magnetized NS ~\cite{Muslimov1992,Rezzolla2001,RezzollaAhmedov2004}. A similar approach is used in studies of compact-star electromagnetic fields in modified gravity: when the metric functions entering the magnetic sector of Maxwell's equations are changed, the magnetic field is modified; when only the gravitomagnetic sector is corrected, the dominant modification appears in the electric field and in the GJ charge density ~\cite{AhmedovMorozova2012,Turimov17,Rayimbaev2021}.
The magnetic field components in the ZAMO frame are given by \cite{Rezzolla2001}

\begin{eqnarray}
        B^{\hat{r}}(r) &=&F(r)\,(\cos\chi\cos\theta + \sin\chi\sin\theta\cos\lambda), \\
    B^{\hat{\theta}}(r) &=&H(r)\,(\cos\chi\sin\theta - \sin\chi\cos\theta\cos\lambda), \\
    B^{\hat{\phi}}(r) &=& G(r)\,(\sin\chi\sin\lambda),
\end{eqnarray}

where $\lambda=\phi-\Omega t$ denotes the rotational phase of the magnetic dipole, $\chi$ is the angle between the magnetic and rotation axes, and the radial functions are adopted from the GR formulation of \cite{Rezzolla2001,Muslimov1992},

\begin{eqnarray}
       F(\eta) &=& \frac{B_0}{\eta^3}\,\frac{f(\eta)}{f(1)},
\\
    H(\eta) &=& G(\eta) = \frac{B_0 N}{\eta^3}
    \left[\frac{3}{2N^2f(1)} - \frac{f(\eta)}{f(1)}\right], 
\end{eqnarray}
where $B_0=2\mu/R^3$ is the Newtonian magnetic field strength at the stellar pole, $\mu$ is the magnetic dipole moment, and $\eta=r/R$. The relativistic correction function $f(\eta)$ is given by ~\cite{Muslimov1992,Rezzolla2001}

\begin{equation}
    f(\eta)= -3\left(\frac{\eta}{\epsilon}\right)^3 \left[
    \frac{\epsilon}{\eta}\left(1+\frac{\epsilon}{2\eta}\right)+\ln\left(1-\frac{\epsilon}{\eta}\right)\right],
\end{equation}
where $\epsilon=2M/R$  is the compactness parameter.

Thus, within the present approximation, the magnetic field retains its GR form and is not directly affected by dCS corrections. The function $f(\eta)$ describes the GR distortion of the dipolar magnetic field due to spacetime curvature. Thus, at the order considered, the magnetic flux surfaces, dipolar field-line geometry, and polar-cap magnetic structure coincide with the GR case. Consequently, the dCS correction does not directly alter the exterior dipolar magnetic field. Its effect on the magnetosphere is mediated by the gravitomagnetic sector, which modifies the frame-dragging frequency and, therefore, the induced electric field, the GJ charge density, and the accelerating electric field.

\subsection{Electric field solution}

The electric field in the magnetosphere of a slowly rotating magnetized NS arises from the relative rotation between the stellar magnetic field and the local inertial frames. In GR, this effect is controlled by the frame-dragging frequency $\omega(r)$ and by the difference between the stellar angular velocity $\Omega$ and the angular velocity of local inertial frames. Therefore, the induced electric field is directly sensitive to the gravitomagnetic structure of spacetime ~\cite{Muslimov1992,Rezzolla2001}. This means that the magnetic geometry remains GR-like, but the corotation condition is modified. Since the GJ charge density and the accelerating electric field depend on the mismatch between $\Omega$ and the local inertial-frame rotation, any change in $\omega(r)$ directly changes the electric field structure.
The electric field components in the ZAMO frame can be written as

\begin{widetext}
\begin{eqnarray}
    E^{\hat{r}} (r,\theta,\phi,t) &=&[f_1(r)+f_3(r)]\cos\chi(3\cos^2\theta-1) +3[g_1(r)+g_3(r)]\sin\chi\sin\theta\cos\theta\cos\lambda ,
\\
    E^{\hat{\theta}}(r,\theta,\phi,t)&=&[f_2(r)+f_4(r)]\cos\chi\sin\theta\cos\theta
    +[g_2(r)+g_4(r)]\sin\chi\sin\lambda
    -[g_5(r)+g_6(r)]\cos2\theta\sin\chi\cos\lambda ,
\\
    E^{\hat{\phi}}(r,\theta,\phi,t)&=&[g_5(r)+g_6(r)]\sin\chi\cos\theta\sin\lambda
    -[g_2(r)+g_4(r)]\sin\chi\cos\theta\sin\lambda .
\end{eqnarray}
\end{widetext}

Here, the radial functions $f_i(r)$ and $g_i(r)$ contain the effects of stellar rotation, spacetime curvature, and frame dragging. The terms proportional to $\Omega$ describe the contribution from the rotation of the stellar magnetic field, while the terms proportional to $\omega^*$ describe the correction associated with the rotation of local inertial frames.

These functions are not independent and satisfy the relations~\cite{Rezzolla2001}
\begin{equation}
    g_1=f_1, \qquad g_3=f_3, \qquad g_5=\frac{f_2}{2}, \qquad g_6=\frac{f_4}{2},
\end{equation}

which reduce the number of independent radial functions entering the solution. The explicit radial functions are 
\begin{eqnarray}
f_1(\eta)&=&\frac{\mu\Omega C^*}{6R^2}\left[\frac{\epsilon^2}{6\eta^2}+\frac{\epsilon}{\eta}+\left(3-\frac{4\eta}{\epsilon}\right)\ln N^2-4\right],
\\
f_2(\eta)&=&\frac{\mu \Omega C^* N}{R^2}\left[\left(1-\frac{2\eta}{\epsilon}\right)\ln N^2 -\frac{\epsilon^2}{6\eta^2 N^2}-2\right],
\\
f_3(\eta)&=&\frac{30 \mu\omega^* \eta^3 C_3}{R^2 \epsilon^5}
\Bigg[\left(\frac{\epsilon^2}{6\eta^2}+\frac{\epsilon}{\eta}-4\right)m
\nonumber\\
&+&\left(3-\frac{4\eta}{\epsilon}\right)\ln N^2 + \frac{\epsilon^2}{10\eta^2}\ln N^2 + \frac{\epsilon^3}{20\eta^3}\Bigg],
\\
f_4(\eta)&=& 
-\frac{180\mu\omega^*\eta^3 N C_3}{ R^2\epsilon^5}
\Bigg[\left(1-\frac{2\eta}{\epsilon}\right)\ln N^2
\nonumber\\
&-&\left(\frac{\epsilon^2}{6\eta^2 N^2} - 2\right)
+ \frac{\epsilon^4}{240\eta^4 N^2}\Bigg],
\\
 g_2(\eta)&=&\frac{3\mu\Omega \eta}{ R^2 \epsilon^3 N^2}\left[\ln N^2+\frac{\epsilon}{\eta}\left(1+\frac{\epsilon}{2\eta}\right)\right] ,
\\
g_4(\eta)&=&\frac{3\mu\omega^* \eta}{ R^2 \epsilon^3 N^2}\left[\ln N^2+\frac{\epsilon}{\eta}\left(1+\frac{\epsilon}{2\eta}\right)\right] ,
\end{eqnarray}
where $C^* = C_1 C_2$ and 
\begin{eqnarray}
    \omega^*=  \frac{\kappa \Omega}{\eta^3}\left[1-\frac{10\pi \hat{\alpha}^2}{\epsilon \eta^3}\left(1+\frac{6\epsilon}{7\eta}+\frac{27\epsilon^2}{40\eta^2}\right)\right],
\end{eqnarray}

and $\hat{\alpha}=\alpha/R^2$ is the dimensionless dCS coupling parameter.

All electric-field expressions reduce to their GR counterparts in the $\hat{\alpha}\rightarrow0$ limit, recovering GR frame-dragging. The constants of integration $C_1$, $C_2$, and $C_3$ are determined from boundary conditions at the stellar surface. These conditions require continuity of the tangential components of the electric field and a discontinuity in the normal component due to surface electric charges. They fix the electric-field configuration and ensure matching between the stellar surface and the exterior magnetosphere. They are given by
\begin{eqnarray}
 C_1&=&-\frac{6}{\epsilon^3}\left[\ln N^2_R+\epsilon\left(1+\frac{\epsilon}{2}\right)\right],
\\
C_2&=&\frac{1}{N^2_R}\left[\left(1-\frac{2}{\epsilon}\right)\ln N^2_R -\frac{\epsilon^2}{6 N^2_R}-2\right]^{-1} ,
\\
C_3&=&\frac{\epsilon^2}{30} C_2 \left[\ln N^2_R + \epsilon \right],
\end{eqnarray}

where $N^2_R=1-\frac{2M}{R}$ at $r=R$.

\section{Goldreich--Julian charge density \label{Sec3}}
The physically relevant quantity for plasma acceleration is the electric field component parallel to the magnetic field, $E_\parallel$. This component determines the extraction and acceleration of charged particles along open magnetic field lines. In the following sections, we use the above electromagnetic field structure to derive the Goldreich--Julian charge density, the electrostatic potential, and the accelerating electric field in the polar-cap region. The GJ charge density determines the charge distribution required to screen the electric field in a rotating NS magnetosphere. In curved spacetime, this quantity is governed by the combined effects of stellar rotation, spacetime curvature, and frame dragging that can be expressed as ~\cite{Muslimov1992,RezzollaAhmedov2004}

\begin{equation}
    \rho_{\rm GJ} = -\frac{1}{4\pi} \nabla \cdot \left[\frac{1}{N}\left(1-\frac{\kappa}{\eta^3}\right)\mathbf{u}\times\mathbf{B}\right],
\end{equation}

where $\mathbf{u}=\boldsymbol{\Omega}\times\mathbf{r}$ is the corotation velocity, $\kappa=\epsilon\beta$, and $\beta=I/I_0$ is the ratio of the stellar moment of inertia to its Newtonian value~\cite{Muslimov1992}.

The factor $1-\kappa/\eta^3$ arises due to frame dragging and represents the deviation of the local inertial-frame rotation from the stellar angular velocity. As a result, the effective corotation condition is modified, thereby changing the charge density required to screen the electric field in the magnetosphere. In the present dCS framework, the dipolar magnetic-field structure remains identical to that of the GR case at leading order with the modified frame-dragging effect. Therefore, the correction to the GJ charge density may originate entirely from the gravitomagnetic, rather than from a change in the magnetic-field geometry. Before deriving the explicit form of the dCS correction, we must examine the geometry of the open magnetic-field-line region, since the GJ charge density plays a dominant role precisely in the polar-cap region.

\subsection{Polar-cap geometry and the last open field line}

Before deriving the GJ charge density in dCS gravity, we must clarify whether the geometry of the open magnetic-field-line region is modified. As discussed above, the diagonal components of the metric remain identical to those of GR, and the dipolar magnetic field coincides with the GR solution for a slowly rotating magnetized NS. Consequently, the magnetic field-line topology is unchanged. This implies that constructing the open-field-line region follows the same procedure as in the GR treatment developed by Muslimov and Tsygan~\cite{Muslimov1992}. Similar approaches have also been adopted in studies of NS magnetospheres within gravity theories, in which the geometry affects magnetic field configurations ~\cite{Rayimbaev2020a,Turimov2019}.

The magnetic flux through a surface bounded by a dipolar field line is given by
\begin{equation} \label{Flux}
    \Psi(\eta)=\frac{\pi R^2 B_0}{\eta}\,\frac{f(\eta)}{f(1)} .
\end{equation}
To describe the area between the last open magnetic field lines, we introduce the dimensionless angular variable $\xi=\theta/\Theta(\eta)$, with $0\leq\xi\leq 1$, which labels individual field lines within the polar-cap region. Here $\Theta(\eta)$ is the polar angle of a given field line at radius $r$, measured relative to the last open field line. The field lines are determined by the flux conservation condition

\begin{equation}
    \Psi(\eta)\sin^2\theta = \Psi(1)\sin^2\theta_0 ,
\end{equation}
where $\theta_0$ is the magnetic colatitude at the stellar surface. Using Eq.~(\ref{Flux}), the equation of a magnetic field line can be written as
\begin{equation} \label{theta}
    \theta = \xi \Theta = \sin^{-1}\left[\sin(\xi \Theta_0)
    \sqrt{\eta\,\frac{f(1)}{f(\eta)}}\right] .
\end{equation}

In the small-angle approximation relevant for the polar-cap region, this reduces to
\begin{equation} \label{Theta}
    \Theta(\eta)=\Theta_0\sqrt{\eta\frac{f(1)}{f(\eta)}} .
\end{equation}

Since the magnetic field structure is unchanged relative to GR, the geometry of open field lines and the polar-cap shape retain their GR forms. In particular, the polar-cap opening angle $\Theta_0$ is set by the last open field line that intersects the light cylinder. For the last open field line, the radius vector satisfies $\theta=\pi/2$ at the light cylinder, which yields
\begin{equation}
    \Theta_0 = \sin^{-1}\left(\frac{R}{R_{\rm LC}}\right)^{1/2}.
\end{equation}
In what follows, we consider the polar-cap region in the small-angle limit $\theta \ll 1$, where we can use the approximations $\sin\theta \simeq \theta$ and $\cos\theta \simeq 1$. Thus, the polar-cap geometry remains identical to that in GR, and deviations in the magnetospheric structure arise not from the magnetic-field configuration but from modified electrodynamics, which we address in the next subsection through the GJ charge density.

%\subsection{General relativistic expression}
The GJ charge density in GR in a rotating spacetime and small angle approximation \cite{Muslimov1992,RezzollaAhmedov2004} takes the form
\begin{equation} \label{rho_gj}\rho_{\rm GJ} = -\frac{\Omega B_0}{2\pi \eta^3 N}\frac{f(\eta)}{f(1)}\Bigg[\left(1-\frac{\kappa}{\eta^3}\right)\cos\chi+ H(\eta)\theta\sin\chi\cos\lambda\Bigg], 
\end{equation}
where the function $H(\eta)$ is given by
\begin{equation}
    H(\eta)=\frac{\epsilon}{\eta}-\frac{\kappa}{\eta^3}
    +\frac{1}{f(\eta)N^2}
    \left(1-\frac{3\epsilon}{2\eta}+\frac{\kappa}{2\eta^3}\right).
\end{equation}

The first term describes the aligned component of the charge density, which frame dragging directly affects through the factor $(1-\kappa/\eta^3)$. The second term represents the leading correction arising from the inclination angle and is important for oblique rotators. This expression forms the basis for the limit of vanishing dCS coupling (GR case).

%\subsection{Chern--Simons correction}

\begin{figure}
\includegraphics[width=8cm,angle=0]{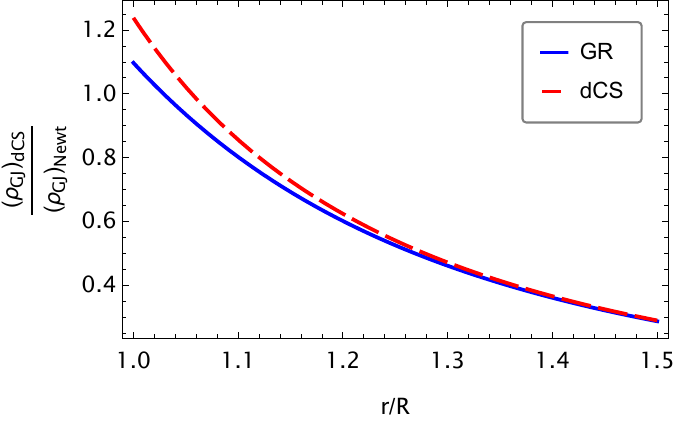}
\caption{Radial dependence of the normalized Goldreich--Julian charge density
$(\rho_{\rm GJ})/(\rho_{\rm GJ})_{\rm Newt}$ as a function of $\eta$ for
$\epsilon=0.4$ and $\chi=0$. The blue curve shows the GR result, while the red dashed curve corresponds to the dCS case with $\hat{\alpha}=0.05$. %The dCS correction increases the required corotation charge density near the stellar surface by modifying the frame-dragging contribution. At larger radii, the correction decays and the dCS profile approaches the GR curve.
}
\label{GJdens}
\end{figure}

In dCS gravity, the GJ charge density is modified by the frame-dragging correction. Since the magnetic-field structure and polar-cap geometry remain unchanged, dCS gravity enters entirely through the modified corotation condition. After algebraic manipulation (see Appendix \ref{AppGJ}), the GJ charge density can be written as
\begin{eqnarray}\nonumber
    \rho_{\rm GJ}=&-&\frac{\Omega B_0}{2\pi \eta^3 N}
    \frac{f(\eta)}{f(1)}\Bigg[\left(1-\frac{\kappa}{\eta^3}W(\eta,\hat{\alpha})\right)\cos\chi 
    \\
    &+&\frac{3}{2}H(\eta)\theta\sin\chi\cos\lambda +{\cal O}(\theta^2) \Bigg],
\end{eqnarray}
  where the function $W(\eta,\hat{\alpha})$ encodes the dCS correction,
\begin{equation}
    W(\eta,\hat{\alpha})=
    1-\frac{10\pi \hat{\alpha}^2}{\epsilon \eta^3}
    \left(1+\frac{6\epsilon}{7\eta}+\frac{27\epsilon^2}{40\eta^2}\right),
\end{equation}

and the modified function $H(\eta)$ can be written as
\begin{equation} \label{Hfunktion}
    H(\eta)=H_{\rm GR}(\eta)+\hat{\alpha}^2H_{\rm dCS}(\eta),
\end{equation}
in which, 
\begin{eqnarray}
H_{\rm dCS}(\eta)&=&\frac{\pi\kappa}{\epsilon\eta^6 }\Bigg[20+\frac{10\epsilon}{\eta}+\frac{46}{7}\frac{\epsilon^2}{\eta^2}-\frac{45}{4}\frac{\epsilon^3}{\eta^3}
\\\nonumber
&-&\frac{1}{N^2 f(\eta)}\left(\frac{20}{3}+\frac{30}{7}\frac{\epsilon}{\eta}+\frac{22}{21}\frac{\epsilon^2}{\eta^2}-\frac{45}{8}\frac{\epsilon^3}{\eta^3}\right)\Bigg]\ ,
\end{eqnarray}
where $H_{\rm dCS}(\eta)$ represents the dCS correction to the oblique part of the GJ charge density. This term enters through the function $H(\eta)$ and therefore affects the contribution proportional to $\theta\sin\chi\cos\phi$. For the aligned case considered in Fig.~\ref{GJdens}, where $\chi=0$, this oblique contribution vanishes, and the radial behavior is controlled by the first term containing $W(\eta,\hat{\alpha})$.

In dCS gravity, the correction reduces the effective frame-dragging effects relative to the GR case. Consequently, the difference between the star's angular velocity and the local inertial-frame rotation becomes larger. Since the GJ charge density quantifies the charge required to maintain corotation, this larger mismatch increases the screening charge density required near the stellar surface.
Fig.~\ref{GJdens} shows this behavior, where the GJ charge density is normalized to its Newtonian value. The blue curve shows the GR case, while the red dashed curve shows the dCS correction for $\hat{\alpha}=0.05$. The GR curve already differs from the Newtonian result due to spacetime curvature and frame-dragging. The upward displacement of the dCS curve relative to the GR curve indicates the additional effect of the CS correction.
The enhancement is strongest close to the stellar surface. This follows from the radial structure of $W(\eta,\hat{\alpha})$, where the correction contains higher inverse powers of $\eta$. Therefore, the dCS contribution is concentrated in the inner magnetosphere. As $\eta$ increases, the correction rapidly decreases and the dCS curve approaches the GR curve. This confirms that the dCS modification primarily alters the near-surface charge density, while the outer magnetosphere remains close to GR behavior.
Thus, the increase in GJ charge density for $\hat{\alpha}\neq0$ has a clear physical origin: the CS correction weakens frame-dragging effects, enhances the effective corotation mismatch, and therefore requires a larger GJ charge density to screen the electric field in the polar-cap region. This interpretation follows the same logic used in GR polar-cap electrodynamics ~\cite{Muslimov1992,RezzollaAhmedov2004}, while the separation between magnetic geometry and gravitomagnetic correction is analogous to the comparative analyses used in modified-gravity magnetosphere studies~\cite{Turimov17,Rayimbaev2021}.

\section{Electrostatic potential and accelerating electric field \label{Sec4}}

The charge density in the open-field-line region can be written as
\begin{eqnarray}
    \rho &=& \frac{\Omega B_0}{2\pi \eta^3 N}\frac{f(\eta)}{f(1)}\left[A(\xi)\cos\chi+\frac{3}{2}D(\xi)\sin\chi\cos\lambda\right].
\end{eqnarray}

%\subsection{Poisson equation in the polar-cap region}

In the small-angle approximation appropriate for the polar-cap region, the Poisson equation for the electrostatic potential takes the form
\begin{eqnarray}\nonumber &&\frac{1}{N^2}\Big\{N\frac{1}{\eta^2}\frac{\partial}{\partial\eta}
    \left(\eta^2\frac{\partial}{\partial\eta}\right)+\frac{1}{N\eta^2} \\&&\left[\frac{1}{\theta}\frac{\partial}{\partial\theta}\left(\theta\frac{\partial}{\partial\theta}\right)+\frac{1}{\theta^2}\frac{\partial^2}{\partial\phi^2}\right]\Big\}\Phi=-4\pi(\rho-\rho_{\rm GJ}).
\end{eqnarray}

Following the standard GR polar-cap formalism of Muslimov and Tsygan, we introduce the dimensionless potential
\begin{equation}
    \Upsilon=\eta\Phi/\Phi_0, \qquad \Phi_0=\Omega R^2B_0,
\end{equation}
together with the angular rescaling $\theta\rightarrow\xi\Theta$. In terms of these variables, the Poisson equation becomes
\begin{eqnarray} \label{Piasson}\nonumber
&&
    \frac{\partial^2\Upsilon}{\partial\eta^2} + \frac{1}{\eta^2N^2\Theta^2}
    \left[\frac{1}{\xi}\frac{\partial}{\partial\xi}\left(\xi \frac{\partial}{\partial\xi} \right)+\frac{1}{\xi^2} 
    \frac{\partial^2}{\partial\phi^2}\right]\Upsilon 
    \\\nonumber
    &=&
    -\frac{2}{\eta^2 N}\frac{f(\eta)}{f(1)}\Big\{\left[1-\frac{\kappa}{\eta^3}W(\eta,\hat{\alpha})+A(\xi)\right]\cos\chi 
   \\& +& \frac{3}{2}\left[\xi\Theta H(\eta) + D(\xi)\right]\sin\chi\cos\lambda\Big\}.
\end{eqnarray}

The angular dependence can then be separated in the form
\begin{eqnarray} \label{Bessel}
\Upsilon(\eta,\xi)=F(\eta,\xi)\cos\chi+S(\eta,\xi)\sin\chi\cos\lambda .
\end{eqnarray}

\subsection{Solution method: mode expansion}
Substituting the above decomposition into the Poisson equation yields two coupled angular-radial equations for the functions $F(\eta,\xi)$ and \(S(\eta,\xi)\),
  \begin{eqnarray}\nonumber 
   && \left[\frac{\partial^2}{\partial\eta^2}
    +\frac{1}{\eta^2N^2\Theta^2}\frac{1}{\xi}
    \frac{\partial}{\partial\xi}\left(\xi\frac{\partial}{\partial\xi}\right)\right]F(\eta,\xi)\\
    &=&
    -\frac{2}{\eta^2N}\frac{f(\eta)}{f(1)} \left[1-\frac{\kappa}{\eta^3}W(\eta,\hat{\alpha})+A(\xi)\right],
    \\ \nonumber
  &&  \left[\frac{\partial^2}{\partial\eta^2} +\frac{1}{\eta^2N\Theta^2} \left(\frac{1}{\xi}\frac{\partial}{\partial\xi}\left(\xi\frac{\partial}{\partial\xi}\right)
    -\frac{1}{\xi^2}\right)\right]S(\eta,\xi)
   \\ &=&
    -\frac{3}{\eta^2N^2}\frac{f(\eta)}{f(1)}
    \left[\xi\Theta H(\eta)+D(\xi)\right].
\end{eqnarray}

To solve these equations, we use a Fourier--Bessel expansion,

\begin{eqnarray}\label{F}
    F(\eta,\xi)=\sum_{n=1}^{\infty}F_n(\eta)J_0(k_n\xi),
\\ \label{S}
        S(\eta,\xi)=\sum_{n=1}^{\infty}S_n(\eta)J_1(w_n\xi),
\end{eqnarray}
where $J_0$ and $J_1$ are Bessel functions, and $k_n$ and $w_n$ denote their corresponding zeros. This mode-expansion procedure follows the standard GR treatment of polar-cap electrodynamics and is also consistent with later modified-gravity extensions of pulsar magnetospheres \cite{Muslimov1992}.
For clarity, we omit the intermediate algebraic reductions here. We collect the orthogonality relations of the Bessel modes, the derivation of the radial equations, and the full analytical steps leading to the near-surface and far-zone solutions in Appendix~\ref{AppFB}.

\subsection{Near-surface solution}

We first consider the region very close to the stellar surface and introduce the dimensionless height above the surface $z=\eta-1\ll1$. In this limit, the radial equations admit exponentially decaying solutions, which satisfy the boundary conditions at the stellar surface. The resulting mode functions can be written as

\begin{widetext}
    \begin{eqnarray}
    F_n &=& 12\kappa\frac{\Theta_0^3 N_1}{k_n^4 J_1(k_n)}
    \left(1-\frac{20\pi \hat{\alpha}^2}{\epsilon}\left(1+\epsilon+\frac{9\epsilon^2}{10}\right)\right) \left[\exp{\left(-\frac{k_n}{\Theta_0 N_1} z\right)}-1+\frac{k_n}{\Theta_0 N_1} z\right],
    \\
    S_n &=& 6\frac{\Theta_0^4 N_1}{w_n^4 J_1(w_n)} H(1)\delta(1) \left[\exp{\left(-\frac{w_n}{\Theta_0 N_1} z\right)}-1 +\frac{w_n}{\Theta_0 N_1} z\right].
\end{eqnarray}

Appendix~D derives these expressions. 
Using the above solutions, the electrostatic potential in the near-surface region and the corresponding accelerating electric field parallel to the magnetic field lines take the form

    \begin{eqnarray}
    \Phi &=& 12 \frac{\Phi_0}{\eta} N_1 \,\kappa\left[1-\frac{20\pi\hat{\alpha}^2}{\epsilon}\left(1+\epsilon+\frac{9\epsilon^2}{10}\right)\right]\Theta_0^3 \cos\chi \sum_{n=1}^{\infty}
    \left[\exp{\left(\frac{k_n(1-\eta)}{\Theta_0 N_1}\right)}-1
    +\frac{k_n(1-\eta)}{\Theta_0 N_1}\right]\frac{J_0(k_n\xi)}{k_n^4J_1(k_n)}
    \nonumber\\
    &+&
    6\frac{\Phi_0}{\eta}N_1 \,\Theta_0^4 H(1)\delta(1)\sin\chi\cos\lambda \sum_{n=1}^{\infty} \left[\exp{\left(\frac{w_n(1-\eta)}{\Theta_0 N_1}\right)}-1
    +\frac{w_n(1-\eta)}{\Theta_0 N_1}\right]\frac{J_1(w_n\xi)}{w_n^4J_2(w_n)} ,
\\
    E_{\parallel} &=& -12 \frac{\Phi_0}{R}\,\kappa\Theta_0^2 \cos\chi\left[1-\frac{20\pi \hat{\alpha}^2}{\epsilon}\left(1+\epsilon+\frac{9\epsilon^2}{10}\right)\right] \sum_{n=1}^{\infty}\left[1-\exp{\left(\frac{k_n(1-\eta)}{\Theta_0 N_1}\right)}\right]\frac{J_0(k_n\xi)}{k_n^3J_1(k_n)}
    \nonumber\\
    &-&
    6\frac{\Phi_0}{R} \Theta_0^3 H(1)\delta(1)\sin\chi\cos\lambda \sum_{n=1}^{\infty} \left[1-\exp{\left(\frac{w_n(1-\eta)}{\Theta_0 N_1}\right)}\right]\frac{J_1(w_n\xi)}{w_n^3J_2(w_n)} .
\end{eqnarray}
\end{widetext}

These expressions describe the initial stage of particle acceleration above the polar cap. The electrostatic potential sets the available potential drop, while $E_{\parallel}$ governs charge extraction and acceleration along the open magnetic field lines. In the present dCS model, the near-surface correction to the GR case arises from the frame-dragging effect and therefore modifies the accelerating electric field without changing the dipolar magnetic field. This role of frame-dragging in establishing the near-surface electrodynamics is fully consistent with the standard GR polar-cap model and with subsequent GR analyses of NS electrodynamics. Modified-gravity studies in our previous related papers \cite{Sayfiyev2025,Rayimbaev2020,Bokhari2021PDU,Rayimbaev15} similarly used the near-surface potential drop and $E_{\parallel}$ as the key quantities controlling primary acceleration and pair creation. 

\subsection{Far-zone solution}

At sufficiently large distances from the stellar surface, the electrostatic potential simplifies and admits an explicit analytical form. Starting from the general solution of the Poisson equation and retaining the leading contributions in the far-zone limit, the electrostatic potential and the corresponding accelerating electric field along the magnetic field lines take the following forms:

\begin{widetext}
    \begin{eqnarray}
    \Phi &=& \frac{\Phi_0 \Theta_0^2}{2}(1-\xi^2)\kappa\left[\left(1-\frac{20\pi \hat{\alpha}^2}{\epsilon}\left(1+\epsilon+\frac{9\epsilon^2}{10}\right)\right)-\frac{1}{\eta^3}\left(1-\frac{10\pi \hat{\alpha}^2}{\epsilon\eta^3}\left(1+\frac{6\epsilon}{7\eta}+\frac{27\epsilon^2}{40\eta^2}\right)\right)\right]\cos\chi
    \nonumber\\ \label{farF}
    &+&\frac{3}{8}\Phi_0\Theta_0^2 \left[\Theta(\eta)H(\eta)-\Theta_0 H(1)\right]\xi(1-\xi^2)\sin\chi\cos\lambda \ ,
\\
    E_{\parallel} &=& -\frac{3}{2}\frac{\Phi_0}{R}\Theta_0^2\kappa \left[\frac{1}{\eta^4}-\frac{20\pi \hat{\alpha}^2}{\epsilon\eta^7}\left(1+\frac{\epsilon}{\eta}+\frac{9\epsilon^2}{10\eta^2}\right)\right](1-\xi^2)\cos\chi -\frac{3}{8}\frac{\Phi_0}{R}\Theta_0^2 \delta(\eta)H(\eta)\xi(1-\xi^2)\sin\chi\cos\lambda \ .
\end{eqnarray}
\end{widetext}

\begin{figure}[ht!]
\includegraphics[width=8cm,angle=0]{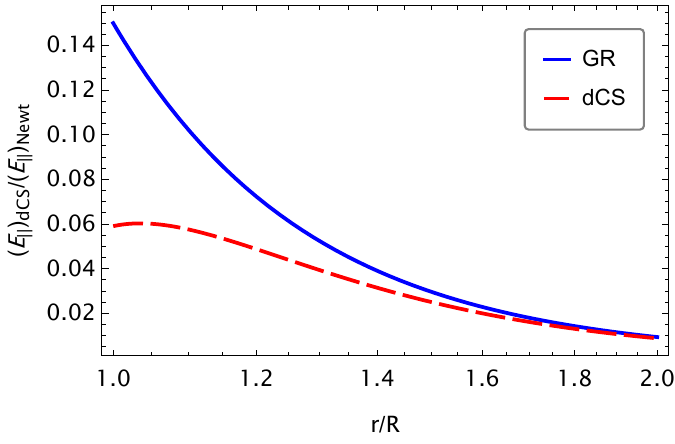}
\caption{Radial dependence of the ratio $(E_{\parallel})_{\rm dCS}/(E_{\parallel})_{\rm Newt}$ as a function of $r/R$ for  the dCS parameter $\hat{\alpha}=0.05$ at $\epsilon=0.4$ and $\chi=0$.}
\label{epar}
\end{figure}

The physical meaning of this solution becomes clear when we analyze the structure of the two contributions. The first term represents the axisymmetric component of the potential drop and is directly controlled by the gravitomagnetic correction encoded in the function $W(\eta,\hat{\alpha})$. The second term describes the oblique contribution associated with the inclination of the magnetic axis and is governed by the functions $\Theta(\eta)$ and $H(\eta)$, which determine the angular structure of the accelerating field.
In order to quantify the impact of the dCS correction on particle acceleration, we plot in Fig.~\ref{epar} the ratio of the accelerating electric field with respect to its Newtonian value for different values of the parameter $\hat{\alpha}$, assuming a fixed compactness $\epsilon=0.4$ and inclination angle $\chi=0$.

As seen from Fig.~\ref{epar}, the accelerating electric field in dCS gravity is systematically lower than in the GR case. The deviation is most pronounced near the stellar surface and increases with $\hat{\alpha}$. As radial distance increases, all curves gradually converge toward the GR profile, indicating that dCS corrections weaken in the outer region of the magnetosphere.
Physically, this behavior implies that dCS modification suppresses particle-acceleration efficiency, particularly in the near-surface region where frame-dragging effects are strongest. Because the accelerating electric field determines the potential drop along open magnetic field lines, this suppression reduces the energy available to primary particles. As a consequence, the conditions for electron--positron pair production become more restrictive, which should shift the pulsar death line toward higher values in the $P$--$\dot{P}$ diagram.
This trend is qualitatively consistent with our previous studies of pulsar electrodynamics in modified gravity. In braneworld and \ae ther gravities \cite{Rayimbaev2019,Bokhari2021PDU}, and in MOG models considered by Rayimbaev \cite{Rayimbaev2020} and collaborators, modifications to the spacetime structure were shown to affect the charge density and accelerating field, shifting the death line. Similarly, in the BBMB geometry analyzed in Ref. \cite{Sayfiyev2025}, changes in near-surface electrodynamics were found to influence particle acceleration and pair-creation thresholds. In contrast, in models where the magnetic field or charge density is strongly modified, the present dCS framework preserves the GR magnetic geometry and introduces corrections primarily through the gravitomagnetic sector. As a result, the deviation from GR manifests mainly through the suppression and radial redistribution of the accelerating electric field rather than through a modification of the field-line structure.

\section{Deathline of radio pulsars in dCS GRAVITY \label{Sec5}}

%\subsection{Pair formation condition}
The radio death line separates pulsars that can sustain an electron--positron pair cascade from those in which pair formation becomes inefficient. In the polar-cap scenario, the electric field accelerates primary electrons along open magnetic field lines, which emit high-energy photons that subsequently convert into secondary electron-- positron pairs in the strong magnetic field. Near the death line, the dominant mechanism for pair production is ICS of thermal photons by relativistic electrons \cite{Muslimov1992,Kantor2004}. The pair cascade can be sustained only if the accelerating potential produces photons above the pair-creation threshold. Otherwise, the cascade ceases, and the pulsar becomes radio quiet.

The death line is determined by the balance between particle acceleration and the ICS pair-production threshold. In dCS gravity, this balance shifts as the accelerating electric field changes.

%\subsection{Particle energy and acceleration}
For the aligned rotator ($\chi=0$), the far-zone potential (\ref{farF}) yields along the magnetic axis 
($\xi\simeq 0$) \cite{Muslimov1992}
\begin{eqnarray}\nonumber
    \Phi(\eta) &\simeq& \frac{\Phi_0\Theta_0^2}{2}\, \kappa
\Bigg[\left(1-\frac{20\pi \hat{\alpha}^2}{\epsilon}\left(1+\epsilon+\frac{9\epsilon^2}{10}\right)\right)
\\
&
-&\frac{1}{\eta^3}\left(1-\frac{10\pi \hat{\alpha}^2}{\epsilon\eta^3}\left(1+\frac{6\epsilon}{7\eta}+\frac{27\epsilon^2}{40\eta^2}\right)\right)
\Bigg].
\end{eqnarray}
The corresponding Lorentz factor of primary electrons is $\gamma=e\Phi/(mc^2)$.
Thus, particle energy depends directly on the accelerating electrostatic potential, which the dCS coupling modifies through frame-dragging effects. Since the dCS correction reduces the magnitude of the accelerating electric field, the available potential drop is reduced, leading to lower particle energies than in the GR case. The reduction in the accelerating potential implies that, at the given distances, pair production becomes less efficient in dCS gravity.
This allows us to formulate the pair-production condition in terms of a radial structure function ~\cite{Muslimov1992,Kantor2004}. 
To express the death line in the $P-\dot P$ diagram, the surface magnetic field is written as
\begin{equation} 
    B_0 \sim \kappa\left(\frac{P}{1{\rm s}}\right)^{7/2} \left(\frac{R}{10{\rm km}}\right)^{-3}10^{12}{\rm G},
\end{equation}
where $B_0$ represents the magnetic field at the star's surface.

%\subsection{Critical condition and death line}
Following the ICS model discussed in Ref. \cite{Kantor2004}, the pair-production condition is

\begin{eqnarray}
    74\left(\frac{B}{10^{12}{\rm G}}\right)^2 P^{-5/2} = {\cal D}(\eta,\epsilon,\hat{\alpha}),
\end{eqnarray}
where the function ${\cal D}(\eta,\epsilon,\hat{\alpha})$ is given by
\begin{eqnarray}\nonumber
    \frac{\eta^2}{ {\cal D}(\eta,\epsilon,\hat{\alpha})}&=&  1-\frac{20\pi \hat{\alpha}^2}{\epsilon}\left(1+\epsilon+\frac{9\epsilon^2}{10}\right) \\&-&\frac{1}{\eta^3}\left[1-\frac{10\pi \hat{\alpha}^2}{\epsilon\eta^3}\left(1+\frac{6\epsilon}{7\eta}+\frac{27\epsilon^2}{40\eta^2}\right)\right].
    \end{eqnarray}

The function ${\cal D}(\eta,\epsilon,\hat{\alpha})$ describes the radial structure of the pair-production condition in the pulsar magnetosphere. The physically relevant point corresponds to the maximum, which determines the maximum altitude at which pair formation is possible. The critical radius $\eta_{\rm cr}$ can be found as a solution of the equation $ \partial_\eta{\cal D}(\eta,\epsilon,\hat{\alpha})=0$, which is, in the GR limit, $\eta_{\rm cr}\simeq1.35$ for typical NSs with mass $M=2$km and radius $R=10$km, in agreement with the classical result of the ICS death-line model~\cite{Kantor2004}.

\begin{figure}[ht!]
\includegraphics[width=8cm,angle=0]{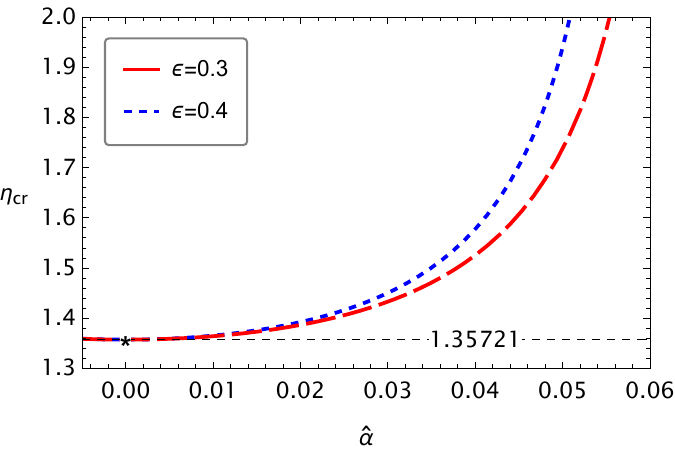}
\caption{Dependence of the critical radius $\eta_{\rm cr}$ on the dCS parameter $\hat{\alpha}$ for different values of the compactness $\epsilon$. The horizontal dashed line indicates the GR value $\eta_{\rm cr}\simeq1.357$. \cite{Kantor2004}}
\label{rmin}
\end{figure}

The critical point $\eta_{\rm cr}$ is determined from ${\cal D}'(\eta)=0$, where ${\cal D}(\eta,\epsilon,\hat{\alpha})$ describes the radial structure of the pair-formation condition derived from the far-zone potential. As shown in Fig.~\ref{rmin}, $\eta_{\rm cr}$ increases with the dCS parameter $\hat{\alpha}$ as a power law for fixed compactness $\epsilon$. The GR limit is recovered at $\hat{\alpha}=0$, where $\eta_{\rm cr}\simeq 1.357$.
Since the dCS correction enters through $\hat{\alpha}^2$, the physical effect depends only on its magnitude; therefore, we express the dependence in terms of $\hat{\alpha}$.
The increase of $\eta_{\rm cr}$ indicates that the pair-formation region shifts to larger distances from the stellar surface. This behavior arises from the reduced accelerating electric field $E_{\parallel}$ and requires higher altitudes to reach the pair-production threshold. In contrast to modified-gravity models such as MOG, braneworld, and conformally coupled gravity scenarios, where both the magnetic-field structure and the effective gravitational potential are altered (~\cite{Rayimbaev2020,Bokhari2021PDU,Sayfiyev2025}), the present dCS model preserves the dipolar magnetic field. As a result, the modification arises solely from the gravitomagnetic effect, thereby explaining the monotonic increase in $\eta_{\rm cr}$.

\begin{figure}[ht!]
\includegraphics[width=8cm,angle=0]{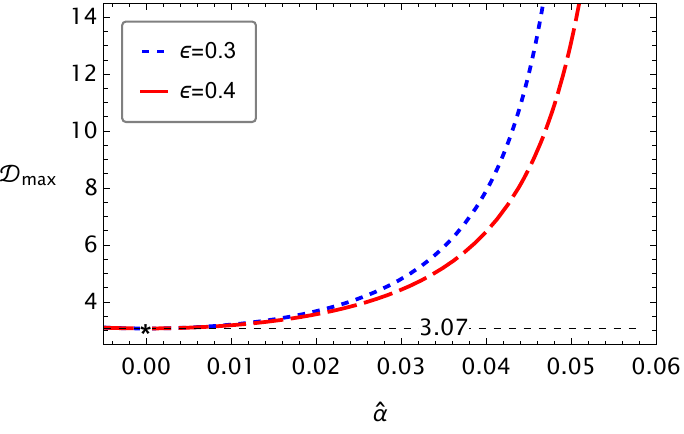}
 \caption{Dependence of the maximum value ${\cal D}_{\rm max}$ on the dCS parameter $\hat{\alpha}$ for different values of $\epsilon$. The dashed line corresponds to the GR value ${\cal D}_{\rm max}\simeq3.07$.}
\label{Dmax}
\end{figure}

The maximum value ${\cal D}_{\max}$ determines the most restrictive point of the pair-production condition and therefore sets the death-line normalization. As shown in Fig.~\ref{Dmax}, ${\cal D}_{\max}$ increases with $\hat{\alpha}$ for all values of $\epsilon$, while the GR limit corresponds to ${\cal D}_{\max}\simeq 3.07$.
A larger value of ${\cal D}_{\max}$ implies that stronger conditions are required to sustain pair production. Physically, this means that either a higher magnetic field or a faster rotation rate is needed to maintain an active pair cascade. This behavior directly reflects the suppression of particle acceleration due to the decrease of the accelerating electric field $E_{\parallel}$. In models where the accelerating field is enhanced, ${\cal D}_{\max}$ typically decreases, leading to more efficient pair production. This opposite trend highlights the distinct role of dCS gravity.
To confront the theoretical predictions with observations, we use two representative pulsar populations: millisecond pulsars (MSPs) and normal (second-period) pulsars. We compile the data from recent observational studies and list them in Appendix~E. Millisecond pulsars are characterized by short periods ($P \sim 10^{-3}$--$10^{-2}$~s) and low spin-down rates \cite{why+25,Kerr2025,bbc+24,bnc+24,vcs+24,prf+24,wpq+24}, while normal pulsars have longer periods ($P \sim 1$--$10$~s) and larger $\dot P$ values \cite{dcm+23,shw+23,wyw+23,psf+22}.

\begin{figure}[ht!]
\includegraphics[width=8cm,angle=0]{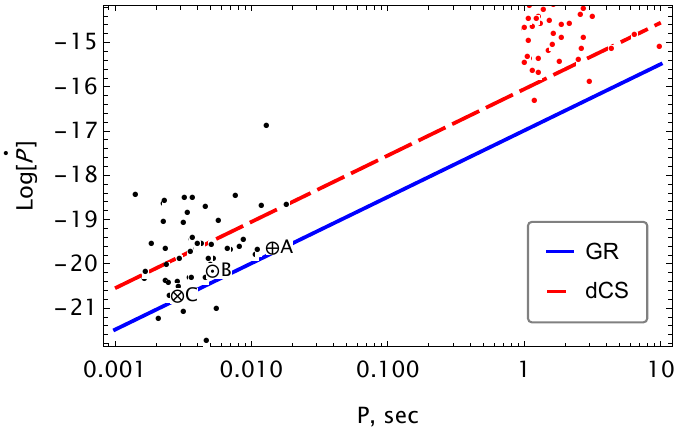}
\caption{Radio pulsar death lines in the $P-\dot P$ diagram for $\epsilon=0.4$. The solid line corresponds to GR, while dashed curves show dCS models for values of $\hat{\alpha}$=0.05.}
\label{Deathline}
\end{figure}
Using these data, we construct the $P$--$\dot P$ diagram shown in Fig.~\ref{Deathline}. Black points denote millisecond pulsars, while red points correspond to normal pulsars. We obtain the theoretical death lines within the dCS framework for different values of $\hat{\alpha}$. The cross-circles A, B, and C mark three representative millisecond pulsars, J~2145$-$0750, J~0024$-$7204~D, and J~0024$-$7204~H, respectively, with their measured $(P,\dot P)$ values taken from the observational data compiled in Appendix~E \cite{why+25,Kerr2025,bbc+24,bnc+24,vcs+24,prf+24,wpq+24}.
In the polar-cap picture, the death line marks the boundary in the $P$--$\dot P$ plane between radio-active pulsars and those unable to sustain pair-creation cascades. Pulsars located above the line are expected to maintain sufficient particle acceleration to trigger pair production, while those below it are less likely to support the cascade required for a plasma-filled magnetosphere and therefore tend toward the radio-quiet regime.
For relatively large values of $\hat{\alpha}$, the predicted death line shifts upward in the $P$--$\dot P$ diagram. At a fixed period, this implies that a larger spin-down rate is required to sustain pair creation and hence radio activity. Physically, this upward shift means the conditions for maintaining an active polar-cap cascade become more restrictive, so the domain of radio-active pulsars shrinks compared with the GR case. In particular, while the agreement with observational data remains reasonable for normal pulsars, likely due to their larger spin-down rates and stronger effective magnetic fields, moderate deviations appear for millisecond pulsars.
Representative examples include the millisecond pulsars J1845+0317 and J1918+0621 \cite{Kerr2025,bbc+24}, and the normal pulsars J1849+0009 and J1903+0851 \cite{wyw+23,psf+22}, which further illustrate the separation between the two populations in the $P$--$\dot P$ diagram.
The upward shift of the death line arises from suppression of the accelerating electric field $E_{\parallel}$, which depends directly on reduced frame dragging. As the accelerating potential drop decreases, primary particles attain lower Lorentz factors, and the generation of high-energy photons responsible for pair creation becomes less efficient. Consequently, the pair-formation threshold is reached only under more extreme conditions, requiring either stronger magnetic fields or larger spin-down rates.
A similar qualitative behavior has been reported in modified-gravity models in Ref.~\cite{Rayimbaev2020} and in the BBMB framework studied in Ref.~\cite{Sayfiyev2025}. However, in those models, the modification often affects both the magnetic-field structure and the charge density. In contrast, in the present framework, the magnetic field remains unchanged, and the modification arises solely through the gravitomagnetic effect. This leads to a clearer physical interpretation of the observed effects. The pulsar data used in constructing the $P$--$\dot P$ diagram are listed in Appendix~E.

\section{Particle acceleration in the polar-cap region in dCS gravity} \label{Sec6}
%\subsection{Equations of motion and electrostatic potential}

The acceleration of charged particles in the polar-cap region is a key mechanism of pulsar emission models. In this region, particles are extracted from the NS surface and accelerated along open magnetic field lines by an electric field parallel to the lines. This field originates from the deviation of the actual space charge density from the GJ charge density.
The motion of a charged particle in curved spacetime is governed by the GR equation of motion \cite{Sakai03}

\begin{eqnarray}
    \frac{d v^{\mu}}{d\tau}+\Gamma^{\mu}_{\nu\lambda}v^{\nu}v^{\lambda} =\frac{e}{m}F^{\mu\nu}v_{\nu} \ .
\end{eqnarray}

This equation consistently describes the dynamics of charged particles near a rotating magnetized NS and forms the basis of the polar-cap acceleration model. In the polar-cap region, where the angular size is small ($\theta\ll1$), the Maxwell and dynamical equations reduce to a one-dimensional form along magnetic field lines \cite{Sakai03,Muslimov1992}. Introducing normalized variables, the Poisson equation for the electrostatic potential can be written as

\begin{eqnarray}
    \frac{N^{2}}{s^2}\frac{d}{ds}
    \left(s^2\frac{d\phi}{ds}\right)-\frac{l(l+1)}{s^2}\phi =
    \frac{B}{B_0 }\left(\frac{j}{V}-\bar{j}\right),
\end{eqnarray}
where the source term is determined by the difference between the actual current density $j$ and the GJ current $\bar{j}$.

The evolution of the Lorentz factor $\gamma$ is governed by

\begin{eqnarray}
    \frac{d}{ds}\left(N\gamma\right)=\frac{1}{N^2}\frac{d\phi}{ds},
\end{eqnarray}
which shows that particle acceleration is directly driven by the electrostatic potential gradient. Here $l$ is the multipole number. The Lorentz factor is defined as
\begin{eqnarray}
    \gamma \equiv -v^\mu u_\mu = \frac{1}{\sqrt{1-V^2}} \ ,
\end{eqnarray}
where $u^\mu$ is the four-velocity of the fiducial observer, and $V$ is the corresponding three-velocity.

Following \cite{Sakai03}, we introduce the normalized variables:
\begin{eqnarray} \label{var}
    j&\equiv &-\frac{2\pi N(s) \rho(s)}{\Omega B(s)}\ ,\qquad \phi(s)\equiv\frac{e}{m}\Phi(s)\ ,\nonumber 
    \\ 
    \bar{j}&\equiv &-\frac{2\pi N(s)\rho_{\rm GJ}(s)}{\Omega B(s)} \ ,\quad s\equiv\sqrt{\frac{2\Omega B_0e}{mc^2}}r \ .\nonumber
\end{eqnarray}
Here $j$ and $\bar{j}$ represent the normalized and GJ current densities, respectively.

Near the NS surface, introducing $y=\eta-1$, the GJ current can be expanded as
\begin{equation}
    \bar{j} \simeq \bar{j}_R \left(1 + \frac{3\omega^*_R}{\Omega} y \right),
\end{equation}
where $\bar{j}_R = 1 - 3\omega^*_R/\Omega$.

In dCS gravity, frame dragging introduces an additional radial dependence in $\bar{j}$. As a result, the balance between $j$ and $\bar{j}$ is modified, directly affecting the accelerating electric field.

%\subsection{Acceleration mechanisms}

The right-hand side of the Poisson equation determines whether acceleration occurs. When the current density deviates from the GJ value ($j\neq\bar{j}$), it generates a non-zero parallel electric field. 

Two regimes are possible:
\begin{itemize}
\item $j<\bar{j}$ (subcritical regime): the electric field changes sign/direction, leading to oscillatory behavior in electron/positron energy. The regime corresponds to the ISC of low-energy photons with high-energy electrons.
\item $j>\bar{j}$ (supercritical regime): a monotonic electric field develops, resulting in continuous acceleration of electrons/positrons through curvature radiation.
\end{itemize}

Gravitational effects modify this balance through frame dragging, which alters $\bar{j}$ and shifts the acceleration condition \cite{Sakai03}. 

In dCS gravity, corrections to frame dragging further modify these effects and enter the expression for $\bar{j}$. As a result, the accelerating electric field and particle energy gain are directly affected.

%\subsection{Numerical results}
\begin{figure}
\includegraphics[width=8cm]{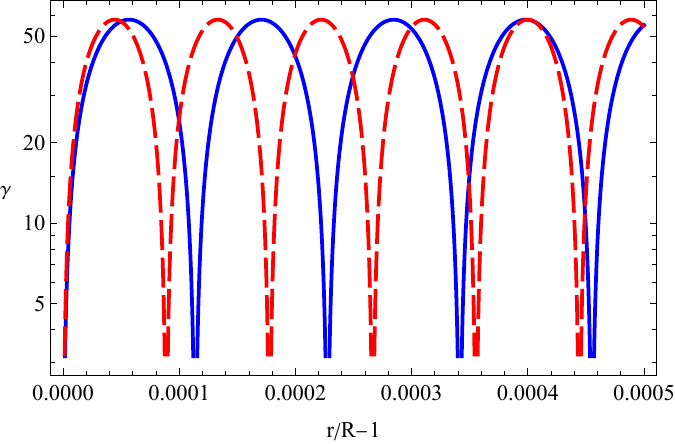}
\includegraphics[width=8cm]{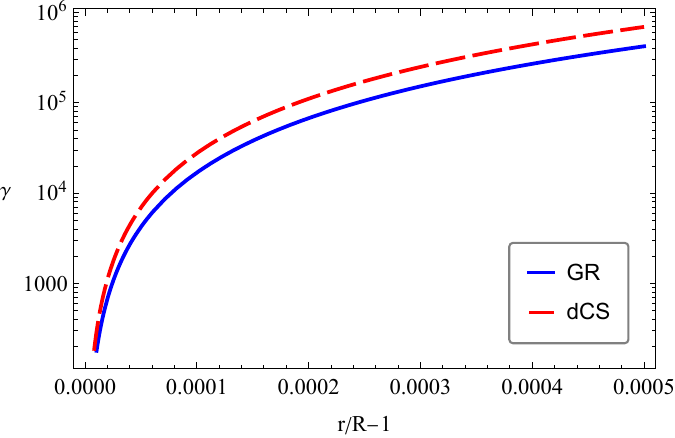}
\caption{Radial dependence of the Lorentz factor $\gamma$ as a function of $(r/R-1)$ for the GR case (solid line) and dCS gravity (red-dashed line). The upper panel shows the subcritical regime for ICS ($j=0.99\bar{j}$), while the lower panel shows the supercritical regime where curvature radiation dominates ($j=1.01\bar{j}$). In the numerical analyses, we chose NS mass as the typical mass $ M=2$ km, radius $ R=10$ km, and the surface electron rest mass $\gamma_R=1.0000001$.}
\label{gamma}
\end{figure}

Figure~\ref{gamma} compares the radial profiles of the Lorentz factor in GR and dCS (with $\hat{\alpha}=0.05$). In the subcritical regime ($j<\bar{j}$), the Lorentz factor oscillates, indicating acceleration and deceleration phases in electron/positron energy.  dCS corrections shift the positions of the maximum and minimum, reflecting the modified charge density and electric-field structure. In the supercritical regime ($j>\bar{j}$), the Lorentz factor increases monotonically with radius. The dCS corrections systematically modify the growth rate of $\gamma$. The results show that charged-particle acceleration in the polar-cap region is highly sensitive to modifications of the spacetime structure. In comparison with the standard GR results of \cite{Muslimov1992} and the extended analysis of \cite{Sakai03}, the dCS corrections introduce additional radial dependence through the modified frame-dragging term. Similar effects were reported in modified gravity studies by \cite{Rayimbaev2020,Turimov2018}, where deviations from GR alter the charge density and accelerating electric field.

\section{Conclusion \label{Summary}}

In this work, we present a consistent analysis of the plasma magnetosphere of a slowly rotating magnetized NS in dCS gravity, focusing on how gravitomagnetic corrections change electromagnetic processes in the polar-cap region.

We have adopted a perturbative dCS spacetime in the slow-rotation and small-coupling regime, where the modification enters exclusively through the off-diagonal metric component. As a result, the metric's diagonal structure and the dipolar magnetic-field configuration remain identical to those in GR. This allows us to isolate the physical impact of frame dragging and to attribute all deviations from the GR case to gravitomagnetic effects.

Within this framework, we derived the GJ charge density and showed that a decrease in the frame-dragging effect increases the effective angular-velocity difference $(\Omega - \omega^*)$. This directly increases the deviation $(\rho - \rho_{GJ})$ near the surface. The ratio of GJ charge density in dCS to that in GR cases exceeds unity near the star surface and gradually approaches the GR limit at larger radii. This confirms that the dCS correction is localized to the inner magnetosphere and is governed by the radial decay in the frame-dragging contribution of dCS gravity.

We then solved the Poisson equation for the electrostatic potential using a mode-expansion method and obtained analytical solutions in both the near-surface and far-zone regions. The resulting expressions show that the accelerating electric field $E_\parallel$ is systematically reduced in the dCS model. The field remains below unity across all radii and decreases as the coupling parameter increases. 

%The simultaneous increase of $(\rho - \rho_{GJ})$ and decrease of $E_\parallel$ is physically consistent. While the charge imbalance is enhanced by reduced frame-dragging, the accelerating field depends on the full solution of the Poisson equation and is controlled by both the source term and the spacetime geometry. As a result, the global structure of the electrostatic potential leads to a net suppression of the accelerating field despite the increased charge deviation.

Using the modified electrostatic potential, we formulated the pair-formation condition within the ICS process and derived the corresponding death-line equation. The analysis shows that the critical radius $\eta_{\rm cr}$ is increased in dCS gravity, indicating that the pair-formation region shifts outward. At the same time, the maximum value of the controlling function $D_{\max}$ increases, implying that stronger conditions are required to sustain pair production.

Together, these effects produce a systematic upward shift of the pulsar death line in the $P$–$\dot P$ diagram. The shift reflects reduced charged-particle acceleration efficiency and the corresponding suppression of pair-production cascades. Comparison with observational data indicates that, for sufficiently large dCS values, the predicted death lines approach the region occupied by millisecond pulsars, while moderate deviations remain for normal pulsars. This suggests that gravitomagnetic corrections can play a measurable role in determining the boundary of radio-loud pulsars.

We have also analyzed particle acceleration in the polar-cap region by solving the equations of motion for charged particles. The results show that the Lorentz-factor evolution is highly sensitive to the modified frame-dragging term. In the subcritical regime, the Lorentz factor oscillates, while in the supercritical regime, the growth rate changes. This demonstrates that even small deviations in frame dragging can significantly affect the energy gain of primary particles.
\appendix
\begin{widetext}
\section{GJ density \label{AppGJ}}
\subsection{Chern--Simons correction}

In dynamical Chern--Simons gravity, the Goldreich--Julian charge density is modified through the corrected frame-dragging term. After straightforward algebraic manipulations, one obtains

    \begin{align}
\rho_{\rm GJ} &= \frac{\Omega B_0}{4\pi \eta^2 f(1)}
\Bigg[\frac{d}{d\eta}\left(\left(1-\frac{\kappa}{\eta^3}W(\eta,\hat{\alpha})\right)
\frac{\eta^2}{2}\frac{d}{d\eta}\left(\frac{f(\eta)}{\eta}\right)\right)\sin\theta\left(\cos\chi\sin\theta - \sin\chi\cos\theta\cos\phi\right)
\nonumber\\&\quad +\left(1-\frac{\kappa}{\eta^3}W(\eta,\hat{\alpha})\right)\frac{f(\eta)}{\eta N}\frac{1}{\sin\theta}\frac{d}{d\theta}\left(\sin^2\theta\left(\cos\chi\cos\theta + \sin\chi\sin\theta\cos\phi\right)\right)\Bigg] \nonumber\\[6pt]
%&= \frac{\Omega B_0}{4\pi \eta^2 f(1)}\Bigg[\frac{d}{d\eta} \left(\left(1-\frac{\kappa}{\eta^3}W(\eta,\hat{\alpha})\right) \frac{\eta^2}{2}\frac{d}{d\eta}\left(\frac{f(\eta)}{\eta}\right)\right)\sin\theta\left(\cos\chi\sin\theta - \sin\chi\cos\theta\cos\phi\right) \nonumber\\&\quad + \left(1-\frac{\kappa}{\eta^3}W(\eta,\hat{\alpha})\right) \frac{f(\eta)}{\eta N}\Bigg(\frac{1}{2}(3\cos2\theta+1)\cos\chi + 3\sin\theta\cos\theta\sin\chi\cos\phi\Bigg)\Bigg]\nonumber\\[6pt]
&\simeq
\frac{\Omega B_0}{2\pi \eta^3 N}\frac{f(\eta)}{f(1)}
\Bigg[\left(1-\frac{\kappa}{\eta^3}W(\eta,\hat{\alpha})\right)\cos\chi\frac{3}{2}
\Bigg(\left(1-\frac{\kappa}{\eta^3}W(\eta,\hat{\alpha})\right)
\nonumber\\
&\qquad
- \frac{\eta N^2}{f(\eta)}\frac{d}{d\eta}
\left(
\left(1-\frac{\kappa}{\eta^3}W(\eta,\hat{\alpha})\right)
\frac{\eta^2}{6}
\frac{d}{d\eta}\left(\frac{f(\eta)}{\eta}\right)
\right)
\Bigg)
\theta\sin\chi\cos\phi
\Bigg]
\nonumber\\[6pt]
&\simeq
\frac{\Omega B_0}{2\pi \eta^3 N}\frac{f(\eta)}{f(1)}
\left[
\left(1-\frac{\kappa}{\eta^3}W(\eta,\hat{\alpha})\right)\cos\chi
+ \frac{3}{2} H(\eta)\,\theta\sin\chi\cos\phi
+ \mathcal{O}(\theta^3)
\right].
\label{ro}
\end{align}
Here
\begin{eqnarray}
    W(\eta,\hat{\alpha})=1-\frac{10\pi \hat{\alpha}^2}{\epsilon \eta^3}\left(1+\frac{6\epsilon}{7\eta}+\frac{27\epsilon^2}{40\eta^2}\right) .
\end{eqnarray}

\section{Fourier--Bessel expansion and radial equations \label{AppFB}}

Substituting Eq.(\ref{Piasson}) into Eqs.(\ref{F}) and (\ref{S}) one can obtain
\begin{eqnarray}
&&\label{Fi}
\sum_{n=1}^{\infty}J_0(k_n \xi)\left[\frac{\partial^2}{\partial\eta^2} - \frac{k_n^2}{\eta^2 N^2\Theta^2}\right]F_n 
=
 -\frac{2}{\eta^2 N^2}\frac{f(\eta)}{f(1)}
\left[1-\frac{\kappa}{\eta^3}W(\eta,\hat{\alpha}) + A(\xi)\right]
\\
&&\label{Si}
\sum_{n=1}^{\infty}J_1(w_n \xi)     \left[\frac{\partial^2}{\partial\eta^2} - 
\frac{w_n^2}{\eta^2 N^2\Theta^2}\right]S_n
=-\frac{3}{\eta^2 N^2}\frac{f(\eta)}{f(1)}\left[\xi \Theta H(\eta) + D(\xi)\right]
\end{eqnarray}

We now multiply by $\xi J_0(k_n\xi)$ both side of the Eq.(\ref{Fi}) and
$\xi J_1(w_n\xi)$ both side of the Eq.(\ref{Si}) integrate over $\xi$, 
and using the condition in Eq.(\ref{Piasson})  

\begin{eqnarray}
&&\label{Fii}
\left[\frac{\partial^2}{\partial\eta^2} - \frac{k_n^2}{\eta^2 N^2\Theta^2}\right]F_n 
=
 -\frac{2}{\eta^2 N^2}\frac{f(\eta)}{f(1)}
\left[\frac{2}{k_n J_1(k_n)}\left(1-\frac{\kappa}{\eta^3}W(\eta,\hat{\alpha})\right) +A_n \right]
\\
&&\label{Sii}
\left[\frac{\partial^2}{\partial\eta^2} - \frac{1}{\eta^2 N^2\Theta^2}\right]S_n
=
 -\frac{3}{\eta^2 N^2}\frac{f(\eta)}{f(1)}\left[\frac{2}{w_n J_2(w_n)}\Theta H(\eta) + D_n\right]
\end{eqnarray}

\subsection{Radial equations and near-surface solution}

Now, let us consider a region very near the surface of a NS
and introduce a dimensionless height above the surface, $z=\eta-1$

\begin{eqnarray}
&&\label{Fz}
\left[
\frac{\partial^2}{\partial z^2}
-
\frac{k_n^2}{\Theta_0^2 N_1^2}
\right]F_n
=
-\frac{2}{N_1^2}
\Bigg[\frac{2}{k_n J_1(k_n)}
\Bigg(1-\kappa W(1,\hat{\alpha})
-3\kappa z\left(1-\frac{20\pi\hat{\alpha}}{\epsilon}\left(1+\epsilon+\frac{9}{10}\epsilon^2\right)\right)
\Bigg)
+
A_n
\Bigg]
\end{eqnarray}
\begin{eqnarray}
    &&\label{Sz}
\left[\frac{\partial^2}{\partial z^2} - \frac{w_n^2}{\Theta^2_0 N_1^2}\right]S_n
=
-\frac{3}{N_1^2}\left[\frac{2}{w_n J_2(w_n)}\theta_0 H(1)(1+\delta(1)\, z)+D_n\right]
\end{eqnarray}

where
$$
\delta(\eta) = \frac{d}{d\eta}\ln \left[H(\eta)\Theta(\eta)\right], 
\qquad N_1^2 = 1-\epsilon
$$

\begin{eqnarray}
&&\label{FF}
\left[\frac{\partial^2}{\partial z^2} - \frac{k_n^2}{\Theta^2_0 N_1^2}\right]F_n 
=
-\frac{2}{N_1^2}
\frac{6\kappa}{k_n J_1(k_n)}\left(1-\frac{20\pi\alpha^2}{\epsilon}\left(1+\epsilon+\frac{9\epsilon^2}{10}\right)\right)
\left(z-\frac{\Theta_0 N_1}{k_n}\right)
\\
&&\label{SS}
\left[\frac{\partial^2}{\partial z^2} - \frac{w_n^2}{\Theta^2_0 N_1^2}\right]S_n
=
-\frac{3}{N_1^2}\frac{2}{w_n J_2(w_n)}
\Theta_0 H(1)\delta(1)\left(z-\frac{\Theta_0 N_1}{w_n}\right)
\end{eqnarray}
Solutions of these equations are
\begin{eqnarray}
\label{FF}
\\\nonumber
&&
F_n = 12\kappa\frac{\Theta_0^3 N_1}{k_n^4 J_1(k_n)}\left(1-\frac{20\pi\alpha^2}{\epsilon}\left(1+\epsilon+\frac{9\epsilon^2}{10}\right)\right)\left[\exp{\left(-\frac{k_n}{\Theta_0 N_1} z\right)}-1+
\frac{k_n}{\Theta_0 N_1} z\right]
\\
&&\label{SS}
S_n = 6\frac{\Theta_0^4 N_1}{w_n^4 J_1(w_n)}
H(1)\delta(1)\left[\exp{\left(-\frac{w_n}{\Theta_0 N_1} z\right)}-1
+\frac{w_n}{\Theta_0 N_1} z\right]
\end{eqnarray}

Small-angle approximation
\begin{eqnarray}
F_1\sin\theta(\sin\theta\cos\chi-\cos\theta\sin\chi\cos\phi)+F_2\left(\frac{1}{2} (3 \cos
2 \theta+1) \cos\chi+3 \sin\theta\cos\theta\sin\chi \cos\phi\right)
\\\nonumber
=2 F_2 \cos\chi-\theta  (F_1-3 F_2) \sin\chi\cos\phi+\theta ^2 
(F_1-3 F_2) \cos\chi+\frac{2}{3} \theta ^3 (F_1-3F_2) \sin\chi\cos\phi+{\cal O}\left(\theta ^4\right)
\end{eqnarray}

\begin{eqnarray}
\label{Ff}
F_n &\simeq & 4\eta\Theta^2_0
\left(1-\frac{\kappa}{\eta^3}
\right)\frac{1}{k_n^3 J_1(k_n)}
\\
\label{Sf}
S_n &\simeq & 6\eta\Theta^2_0
\left[\Theta(\eta)H(\eta)-\Theta_0 H(1) 
\right]\frac{1}{w_n^3 J_2(w_n)}
\end{eqnarray}

\begin{eqnarray}
\Upsilon(\eta,\xi) &=& \sum_{n=1}^{\infty} F_n(\eta)J_0(k_n \xi)\cos\chi 
+ \sum_{n=1}^{\infty} S_n(\eta)J_1(w_n \xi)\sin\chi\cos\phi 
\end{eqnarray}

\begin{eqnarray}
\Phi &=& \frac{1}{2} \Phi_0
\Theta_0^2 
\left(1-\frac{\kappa}{\eta^3}\right)
(1-\xi^2)\cos\chi 
+
\frac{3}{8}\Phi_0\Theta_0^2 
\left[\Theta(\eta)H(\eta)-\Theta_0 H(1) \right]\xi(1-\xi^2)\sin\chi\cos\phi\ ,
\\\nonumber
E &=& -\frac{3}{2} \frac{\Phi_0}{R}
\Theta_0^2
\frac{\kappa}{\eta^4}(1-\xi^2) \cos\chi -
\frac{3}{8}\frac{\Phi_0}{R}\Theta_0^2 
\delta(\eta)H(\eta)\xi(1-\xi^2)\sin\chi\cos\phi\ ,
\end{eqnarray}
\end{widetext}

\bibliographystyle{apsrev4-1}  
\bibliography{references}

@article{Muslimov1992,
  author  = {Muslimov, A. G. and Tsygan, A. I.},
  title   = {General relativistic electric potential drops above pulsar polar caps},
  journal = {Monthly Notices of the Royal Astronomical Society},
  volume  = {255},
  pages   = {61--70},
  year    = {1992},
  doi     = {10.1093/mnras/255.1.61}
}

@article{Turimov2021,
  author  = {Turimov, Bobur and Stuchl{\'\i}k, Zden{\v{e}}k and Rayimbaev, Javlon and Abdujabbarov, Ahmadjon},
  title   = {General relativistic effects in neutron star electrodynamics},
  journal = {Physical Review D},
  volume  = {103},
  number  = {12},
  pages   = {124039},
  year    = {2021},
  doi     = {10.1103/PhysRevD.103.124039},
  eprint  = {2106.08337},
  archivePrefix = {arXiv},
  primaryClass = {gr-qc}
}

@article{Goldreich1969,
  author  = {Goldreich, Peter and Julian, William H.},
  title   = {Pulsar Electrodynamics},
  journal = {The Astrophysical Journal},
  volume  = {157},
  pages   = {869--880},
  year    = {1969},
  doi     = {10.1086/150119}
}

@article{Harding1998,
  author  = {Harding, Alice K. and Muslimov, Alexander G.},
  title   = {Particle Acceleration Zones above Pulsar Polar Caps: Electron and Positron Pair Formation Fronts},
  journal = {The Astrophysical Journal},
  volume  = {508},
  pages   = {328--346},
  year    = {1998},
  doi     = {10.1086/306394},
  eprint  = {astro-ph/9805132},
  archivePrefix = {arXiv}
}

@article{Harding2002,
  author  = {Harding, Alice K. and Muslimov, Alexander G. and Zhang, Bing},
  title   = {Regimes of Pulsar Pair Formation and Particle Energetics},
  journal = {The Astrophysical Journal},
  volume  = {576},
  pages   = {366--375},
  year    = {2002},
  doi     = {10.1086/341633},
  eprint  = {astro-ph/0205077},
  archivePrefix = {arXiv}
}

@article{Pacini68,
  author  = {Pacini, Franco},
  title   = {Rotating Neutron Stars, Pulsars and Supernova Remnants},
  journal = {Nature},
  volume  = {219},
  pages   = {145--146},
  year    = {1968},
  doi     = {10.1038/219145a0}
}

@article{Gold68,
  author  = {Gold, T.},
  title   = {Rotating Neutron Stars as the Origin of the Pulsating Radio Sources},
  journal = {Nature},
  volume  = {218},
  pages   = {731--732},
  year    = {1968},
  doi     = {10.1038/218731a0}
}

@article{JackiwPi2003,
  author  = {Jackiw, R. and Pi, S.-Y.},
  title   = {Chern-Simons modification of general relativity},
  journal = {Physical Review D},
  volume  = {68},
  pages   = {104012},
  year    = {2003},
  doi     = {10.1103/PhysRevD.68.104012},
  eprint  = {gr-qc/0308071},
  archivePrefix = {arXiv}
}

@article{AlexanderYunes2009,
  author  = {Alexander, Stephon and Yunes, Nicolas},
  title   = {Chern-Simons modified general relativity},
  journal = {Physics Reports},
  volume  = {480},
  pages   = {1--55},
  year    = {2009},
  doi     = {10.1016/j.physrep.2009.07.002},
  eprint  = {0907.2562},
  archivePrefix = {arXiv}
}

@article{GrumillerYunes2008,
  author  = {Grumiller, Daniel and Yunes, Nicolas},
  title   = {How do black holes spin in Chern-Simons modified gravity?},
  journal = {Physical Review D},
  volume  = {77},
  pages   = {044015},
  year    = {2008},
  doi     = {10.1103/PhysRevD.77.044015},
  eprint  = {0711.1868},
  archivePrefix = {arXiv}
}

@article{YunesPretorius2009,
  author  = {Yunes, Nicolas and Pretorius, Frans},
  title   = {Dynamical Chern-Simons modified gravity. I. Spinning black holes in the slow-rotation approximation},
  journal = {Physical Review D},
  volume  = {79},
  pages   = {084043},
  year    = {2009},
  doi     = {10.1103/PhysRevD.79.084043},
  eprint  = {0902.4669},
  archivePrefix = {arXiv}
}

@article{KonnoMatsuyamaTanda2009,
  author  = {Konno, Kohkichi and Matsuyama, Toyoki and Tanda, Satoshi},
  title   = {Rotating Black Hole in Extended Chern-Simons Modified Gravity},
  journal = {Progress of Theoretical Physics},
  volume  = {122},
  pages   = {561--568},
  year    = {2009},
  doi     = {10.1143/PTP.122.561},
  eprint  = {0902.4767},
  archivePrefix = {arXiv}
}

@article{CardosoGualtieri2009,
  author  = {Cardoso, Vitor and Gualtieri, Leonardo},
  title   = {Perturbations of Schwarzschild black holes in dynamical Chern-Simons modified gravity},
  journal = {Physical Review D},
  volume  = {80},
  pages   = {064008},
  year    = {2009},
  doi     = {10.1103/PhysRevD.80.064008},
  eprint  = {0907.5008},
  archivePrefix = {arXiv}
}

@article{MolinaPaniCardosoGualtieri2010,
  author  = {Molina, C. and Pani, Paolo and Cardoso, Vitor and Gualtieri, Leonardo},
  title   = {Gravitational signature of Schwarzschild black holes in dynamical Chern-Simons gravity},
  journal = {Physical Review D},
  volume  = {81},
  pages   = {124021},
  year    = {2010},
  doi     = {10.1103/PhysRevD.81.124021},
  eprint  = {1004.4007},
  archivePrefix = {arXiv}
}

@article{YagiYunesTanaka2012,
  author  = {Yagi, Kent and Yunes, Nicolas and Tanaka, Takahiro},
  title   = {Slowly rotating black holes in dynamical Chern-Simons gravity: Deformation quadratic in the spin},
  journal = {Physical Review D},
  volume  = {86},
  pages   = {044037},
  year    = {2012},
  doi     = {10.1103/PhysRevD.86.044037},
  eprint  = {1206.6130},
  archivePrefix = {arXiv}
}

@article{Stein2014,
  author  = {Stein, Leo C.},
  title   = {Rapidly rotating black holes in dynamical Chern-Simons gravity: Decoupling limit solutions and breakdown},
  journal = {Physical Review D},
  volume  = {90},
  pages   = {044061},
  year    = {2014},
  doi     = {10.1103/PhysRevD.90.044061},
  eprint  = {1407.2350},
  archivePrefix = {arXiv}
}

@article{McNeesSteinYunes2016,
  author  = {McNees, Robert and Stein, Leo C. and Yunes, Nicol{\'a}s},
  title   = {Extremal black holes in dynamical Chern-Simons gravity},
  journal = {Classical and Quantum Gravity},
  volume  = {33},
  number  = {23},
  pages   = {235013},
  year    = {2016},
  doi     = {10.1088/0264-9381/33/23/235013},
  eprint  = {1512.05453},
  archivePrefix = {arXiv}
}

@article{DelsateHerdeiroRadu2018,
  author  = {Delsate, Terence and Herdeiro, Carlos and Radu, Eugen},
  title   = {Non-perturbative spinning black holes in dynamical Chern-Simons gravity},
  journal = {Physics Letters B},
  volume  = {787},
  pages   = {8--15},
  year    = {2018},
  doi     = {10.1016/j.physletb.2018.09.060},
  eprint  = {1806.06700},
  archivePrefix = {arXiv}
}

@article{OkounkovaSteinScheelTeukolsky2019,
  author  = {Okounkova, Maria and Stein, Leo C. and Scheel, Mark A. and Teukolsky, Saul A.},
  title   = {Numerical binary black hole collisions in dynamical Chern-Simons gravity},
  journal = {Physical Review D},
  volume  = {100},
  pages   = {104026},
  year    = {2019},
  doi     = {10.1103/PhysRevD.100.104026},
  eprint  = {1906.08789},
  archivePrefix = {arXiv}
}

@article{YunesPsaltisOzelLoeb2010,
  author  = {Yunes, Nicolas and Psaltis, Dimitrios and {\"O}zel, Feryal and Loeb, Abraham},
  title   = {Constraining parity violation in gravity with measurements of neutron-star moments of inertia},
  journal = {Physical Review D},
  volume  = {81},
  pages   = {064020},
  year    = {2010},
  doi     = {10.1103/PhysRevD.81.064020},
  eprint  = {0912.2736},
  archivePrefix = {arXiv}
}

@article{AliHaimoudChen2011,
  author  = {Ali-Ha{\"i}moud, Yacine and Chen, Yanbei},
  title   = {Slowly rotating stars in dynamical Chern-Simons gravity},
  journal = {Physical Review D},
  volume  = {84},
  pages   = {124033},
  year    = {2011},
  doi     = {10.1103/PhysRevD.84.124033},
  eprint  = {1110.5329},
  archivePrefix = {arXiv}
}

@article{YagiSteinYunesTanaka2013,
  author  = {Yagi, Kent and Stein, Leo C. and Yunes, Nicolas and Tanaka, Takahiro},
  title   = {Isolated and binary neutron stars in dynamical Chern-Simons gravity},
  journal = {Physical Review D},
  volume  = {87},
  pages   = {084058},
  year    = {2013},
  doi     = {10.1103/PhysRevD.87.084058},
  eprint  = {1302.1918},
  archivePrefix = {arXiv}
}

@article{YagiYunes2013,
  author  = {Yagi, Kent and Yunes, Nicolas and Tanaka, Takahiro},
  title   = {Slowly Rotating Neutron Stars in Dynamical Chern-Simons Gravity},
  journal = {Physical Review D},
  volume  = {86},
  pages   = {044037},
  year    = {2012},
  doi     = {10.1103/PhysRevD.86.044037},
  eprint  = {1206.6130},
  archivePrefix = {arXiv},
  note    = {Same paper as YagiYunesTanaka2012}
}

@article{GuptaMajumderYagiYunes2018,
  author  = {Gupta, Toral and Majumder, Barun and Yagi, Kent and Yunes, Nicol{\'a}s},
  title   = {I-Love-Q relations for neutron stars in dynamical Chern-Simons gravity},
  journal = {Classical and Quantum Gravity},
  volume  = {35},
  number  = {2},
  pages   = {025009},
  year    = {2018},
  doi     = {10.1088/1361-6382/aa9c68},
  eprint  = {1709.09174},
  archivePrefix = {arXiv}
}

@article{Rezzolla01c,
  author  = {Rezzolla, L. and Ahmedov, B. J. and Miller, J. C.},
  title   = {General relativistic electromagnetic fields of a slowly rotating magnetized neutron star -- I. Formulation of the equations},
  journal = {Monthly Notices of the Royal Astronomical Society},
  volume  = {322},
  pages   = {723--740},
  year    = {2001},
  doi     = {10.1046/j.1365-8711.2001.04161.x},
  eprint  = {astro-ph/0011316},
  archivePrefix = {arXiv}
}

@article{Rezzolla01d,
  author  = {Rezzolla, L. and Ahmedov, B. J. and Miller, J. C.},
  title   = {Stationary Electromagnetic Fields of a Slowly Rotating Magnetized Neutron Star in General Relativity},
  journal = {Foundations of Physics},
  volume  = {31},
  pages   = {1051--1065},
  year    = {2001},
  doi     = {10.1023/A:1017574223222},
  eprint  = {gr-qc/0108057},
  archivePrefix = {arXiv}
}

@article{Rezzolla2001,
  author  = {Rezzolla, Luciano and Ahmedov, Bobomurat J. and Miller, Jonas C.},
  title   = {General relativistic electromagnetic fields of a slowly rotating magnetized neutron star},
  journal = {Monthly Notices of the Royal Astronomical Society},
  volume  = {322},
  pages   = {723--740},
  year    = {2001},
  doi     = {10.1046/j.1365-8711.2001.04161.x},
  eprint  = {astro-ph/0011316},
  archivePrefix = {arXiv}
}

@article{Fattoyev2008PhRvD,
  author  = {Ahmedov, B. J. and Fattoyev, F. J.},
  title   = {Magnetic fields of spherical compact stars in a braneworld},
  journal = {Physical Review D},
  volume  = {78},
  pages   = {047501},
  year    = {2008},
  doi     = {10.1103/PhysRevD.78.047501},
  eprint  = {gr-qc/0608039},
  archivePrefix = {arXiv}
}

@article{Turimov17,
  author  = {Turimov, Bobur and Ahmedov, Bobomurat and Abdujabbarov, Ahmadjon and Bambi, Cosimo},
  title   = {Electromagnetic fields of slowly rotating magnetized compact stars in braneworld gravity},
  journal = {Physical Review D},
  volume  = {95},
  pages   = {084037},
  year    = {2017},
  doi     = {10.1103/PhysRevD.95.084037},
  eprint  = {1702.08277},
  archivePrefix = {arXiv}
}

@article{Turimov18a,
  author  = {Turimov, B.},
  title   = {Electromagnetic fields in vicinity of tidal charged static black hole},
  journal = {International Journal of Modern Physics D},
  volume  = {27},
  pages   = {1850092},
  year    = {2018},
  doi     = {10.1142/S021827181850092X}
}

@article{Rayimbaev2020,
  author  = {Rayimbaev, Javlon and Tadjimuratov, Pulat},
  title   = {Can modified gravity silence radio-loud pulsars?},
  journal = {Physical Review D},
  volume  = {102},
  pages   = {024019},
  year    = {2020},
  doi     = {10.1103/PhysRevD.102.024019}
}

@article{Rayimbaev2020b,
  author  = {Rayimbaev, Javlon and Tadjimuratov, Pulat},
  title   = {Can modified gravity silence radio-loud pulsars?},
  journal = {Physical Review D},
  volume  = {102},
  pages   = {024019},
  year    = {2020},
  doi     = {10.1103/PhysRevD.102.024019}
}

@article{Sakai03,
  author  = {Sakai, Nobuyuki and Shibata, Shinpei},
  title   = {General Relativistic Electromagnetism and Particle Acceleration in a Pulsar Polar Cap},
  journal = {The Astrophysical Journal},
  volume  = {584},
  pages   = {427--432},
  year    = {2003},
  doi     = {10.1086/345602}
}

@article{Rayimbaev2019IJMPD,
  author  = {Rayimbaev, Javlon and Turimov, Bobur and Palvanov, Satimbay},
  title   = {Plasma magnetosphere of slowly rotating magnetized neutron star in braneworld},
  journal = {International Journal of Modern Physics: Conference Series},
  volume  = {49},
  pages   = {1960019},
  year    = {2019},
  doi     = {10.1142/S201019451960019X}
}

@article{Cheng1986a,
  author  = {Cheng, K. S. and Ho, C. and Ruderman, M.},
  title   = {Energetic radiation from rapidly spinning pulsars. I - Outer magnetosphere gaps},
  journal = {The Astrophysical Journal},
  volume  = {300},
  pages   = {500},
  year    = {1986},
  doi     = {10.1086/163829}
}

@article{Cheng1986b,
  author  = {Cheng, K. S. and Ho, C. and Ruderman, M.},
  title   = {Energetic radiation from rapidly spinning pulsars. II - Outer gap acceleration},
  journal = {The Astrophysical Journal},
  volume  = {300},
  pages   = {522},
  year    = {1986},
  doi     = {10.1086/163830}
}

@article{Daugherty1982,
  author  = {Daugherty, J. K. and Harding, A. K.},
  title   = {Electromagnetic cascades in pulsars},
  journal = {The Astrophysical Journal},
  volume  = {252},
  pages   = {337},
  year    = {1982},
  doi     = {10.1086/159561}
}

@article{Daugherty1996,
  author  = {Daugherty, J. K. and Harding, A. K.},
  title   = {Gamma-Ray Pulsars: Emission from Extended Polar Cap Cascades},
  journal = {The Astrophysical Journal},
  volume  = {458},
  pages   = {278},
  year    = {1996},
  doi     = {10.1086/176808}
}

@article{Ruderman75,
  author  = {Ruderman, M. A. and Sutherland, P. G.},
  title   = {Theory of pulsars: polar gaps, sparks, and coherent microwave radiation},
  journal = {The Astrophysical Journal},
  volume  = {196},
  pages   = {51},
  year    = {1975},
  doi     = {10.1086/153393}
}

@article{Chen93,
  author  = {Chen, K. and Ruderman, M.},
  title   = {Pulsar death lines and death valley},
  journal = {The Astrophysical Journal},
  volume  = {402},
  pages   = {264},
  year    = {1993},
  doi     = {10.1086/172129}
}

@article{Kantor04,
  author  = {Kantor, E. M. and Tsygan, A. I.},
  title   = {Death lines of radio pulsars in the inverse Compton scattering model},
  journal = {Astronomy Reports},
  volume  = {48},
  pages   = {1029},
  year    = {2004},
  doi     = {10.1134/1.1829290}
}

@article{Kantor2004,
  author  = {Kantor, E. M. and Tsygan, A. I.},
  title   = {Death lines of radio pulsars in the inverse Compton scattering model},
  journal = {Astronomy Reports},
  volume  = {48},
  pages   = {1029},
  year    = {2004},
  doi     = {10.1134/1.1829290}
}

@article{Morozova12,
  author  = {Morozova, V. S. and Ahmedov, B. J. and Zanotti, O.},
  title   = {Explaining the subpulse drift velocity of pulsar magnetosphere within the space-charge limited flow model},
  journal = {Monthly Notices of the Royal Astronomical Society},
  volume  = {444},
  pages   = {1144},
  year    = {2014},
  doi     = {10.1093/mnras/stu1486}
}

@article{Zhang96a,
  author  = {Zhang, B. and Harding, A. K. and Muslimov, A. G.},
  title   = {Radio Pulsar Death Line Revisited: Is PSR J2144-3933 Anomalous?},
  journal = {The Astrophysical Journal Letters},
  volume  = {531},
  pages   = {L135},
  year    = {2000},
  doi     = {10.1086/312542}
}

@article{AhmedovMorozova2012,
  author  = {Morozova, V. S. and Ahmedov, B. J.},
  title   = {Electromagnetic fields of slowly rotating compact magnetized stars in braneworld},
  journal = {Astrophysics and Space Science},
  volume  = {333},
  pages   = {133--142},
  year    = {2011},
  doi     = {10.1007/s10509-010-0560-2},
  eprint  = {1012.2190},
  archivePrefix = {arXiv}
}

@article{Ahmedov2012,
  author  = {Morozova, V. S. and Ahmedov, B. J.},
  title   = {Electromagnetic fields of slowly rotating compact magnetized stars in braneworld},
  journal = {Astrophysics and Space Science},
  volume  = {333},
  pages   = {133--142},
  year    = {2011},
  doi     = {10.1007/s10509-010-0560-2},
  eprint  = {1012.2190},
  archivePrefix = {arXiv}
}

@article{Rayimbaev2019,
  author  = {Rayimbaev, Javlon and Turimov, Bobur and Palvanov, Satimbay},
  title   = {Plasma magnetosphere of slowly rotating magnetized neutron star in braneworld},
  journal = {International Journal of Modern Physics: Conference Series},
  volume  = {49},
  pages   = {1960019},
  year    = {2019},
  doi     = {10.1142/S201019451960019X}
}

@article{Sayfiyev2025,
  author  = {Sayfiyev, Sherzod and Bokhari, Ashfaque H. and Ahmedov, Bobomurat and Rayimbaev, Javlon},
  title   = {Vacuum and plasma magnetosphere around rotating magnetized neutron stars in Bocharova--Bronnikov--Melnikov--Bekenstein geometry},
  journal = {The European Physical Journal C},
  volume  = {85},
  pages   = {1345},
  year    = {2025},
  doi     = {10.1140/epjc/s10052-025-14899-z}
}

@article{Bokhari2021PDU,
  author  = {Bokhari, Ashfaque Hussain and Rayimbaev, Javlon and Ahmedov, Bobomurat},
  title   = {Radio loudness and spindown of pulsars in Einstein-aether gravity},
  journal = {Physics of the Dark Universe},
  volume  = {34},
  pages   = {100901},
  year    = {2021},
  doi     = {10.1016/j.dark.2021.100901}
}

@article{Hartle1967,
  author  = {Hartle, James B.},
  title   = {Slowly Rotating Relativistic Stars. I. Equations of Structure},
  journal = {The Astrophysical Journal},
  volume  = {150},
  pages   = {1005},
  year    = {1967},
  doi     = {10.1086/149400}
}

@article{Juraeva2022ArabJMath,
  author  = {Juraeva, Nozima and Rayimbaev, Javlon and Haydarov, Kamoliddin and Umaraliyev, Maksud and Abdujabbarov, Ahmadjon},
  title   = {Constraining spacetime deformation based on astrophysical observations from radio pulsars},
  journal = {Arabian Journal of Mathematics},
  volume  = {11},
  pages   = {133--139},
  year    = {2022},
  doi     = {10.1007/s40065-022-00370-4}
}

@article{Rayimbaev2021,
  author  = {Rayimbaev, Javlon and Turimov, Bobur and Ahmedov, Bobomurat},
  title   = {Magnetized particle motion and electromagnetic fields around compact objects in modified gravity},
  journal = {The European Physical Journal C},
  volume  = {81},
  pages   = {1022},
  year    = {2021},
  doi     = {10.1140/epjc/s10052-021-09807-4}
}

@article{Turimov2018,
  author  = {Turimov, B. S. and Ahmedov, B. J. and Abdujabbarov, A. A.},
  title   = {Electromagnetic fields of slowly rotating magnetized neutron stars in modified gravity},
  journal = {Physical Review D},
  volume  = {98},
  pages   = {084039},
  year    = {2018},
  doi     = {10.1103/PhysRevD.98.084039}
}

@article{Turimov2019,
  author  = {Turimov, Boburjon and Ahmedov, Bobomurat and Abdujabbarov, Ahmadjon},
  title   = {Electromagnetic fields and particle acceleration around compact objects in modified gravity},
  journal = {Physical Review D},
  volume  = {100},
  pages   = {084038},
  year    = {2019},
  doi     = {10.1103/PhysRevD.100.084038}
}

@article{Rayimbaev2020a,
  author  = {Rayimbaev, Javlon and Abdujabbarov, Ahmadjon and Ahmedov, Bobomurat},
  title   = {Magnetized particle motion and electromagnetic processes around compact objects in modified gravity},
  journal = {Physical Review D},
  volume  = {101},
  pages   = {104033},
  year    = {2020},
  doi     = {10.1103/PhysRevD.101.104033}
}

@article{Rayimbaev15,
  author  = {Rayimbaev, J. R.},
  title   = {Magnetized particle motion around non-Schwarzschild black hole immersed in an external uniform magnetic field},
  journal = {Astrophysics and Space Science},
  volume  = {361},
  pages   = {288},
  year    = {2016},
  doi     = {10.1007/s10509-016-2879-9}
}

@article{why+25,
  author  = {{Wang}, P. F. and {Han}, J. L. and {Yang}, Z. L. and others},
  title   = {{The FAST Galactic Plane Pulsar Snapshot Survey. VIII. 116 Binary Pulsars}},
  journal = {Research in Astronomy and Astrophysics},
  volume  = {25},
  pages   = {014003},
  year    = {2025},
  doi     = {10.1088/1674-4527/ada3b8},
  eprint  = {2412.03062},
  archivePrefix = {arXiv}
}

@article{Kerr2025,
  author  = {Kerr, Matthew and Johnston, Simon and Clark, Colin J. and others},
  title   = {Discovery and Timing of Four Gamma-Ray Millisecond Pulsars},
  journal = {The Astrophysical Journal},
  volume  = {984},
  number  = {2},
  pages   = {180},
  year    = {2025},
  doi     = {10.3847/1538-4357/adc7a6},
  eprint  = {2503.12636},
  archivePrefix = {arXiv}
}

@article{bbc+24,
  author  = {{Bangale}, P. and {Bhattacharyya}, B. and {Camilo}, F. and others},
  title   = {A 350 MHz Green Bank Telescope Survey of Unassociated Fermi LAT Sources: Discovery and Timing of 10 Millisecond Pulsars},
  journal = {The Astrophysical Journal},
  volume  = {966},
  pages   = {161},
  year    = {2024},
  doi     = {10.3847/1538-4357/ad2994}
}

@article{bnc+24,
  author  = {{Burgay}, M. and {Nieder}, L. and {Clark}, C. J. and others},
  title   = {Radio and gamma-ray timing of TRAPUM L-band Fermi pulsar survey discoveries},
  journal = {Astronomy \& Astrophysics},
  volume  = {691},
  pages   = {A315},
  year    = {2024},
  doi     = {10.1051/0004-6361/202451530}
}

@article{vcs+24,
  author  = {{Vleeschower}, L. and {Corongiu}, A. and {Stappers}, B. W. and others},
  title   = {Discoveries and timing of pulsars in M62},
  journal = {Monthly Notices of the Royal Astronomical Society},
  volume  = {530},
  pages   = {1436--1456},
  year    = {2024},
  doi     = {10.1093/mnras/stae816}
}

@article{prf+24,
  author  = {{Padmanabh}, P. V. and {Ransom}, S. M. and {Freire}, P. C. C. and others},
  title   = {Discovery and timing of ten new millisecond pulsars in the globular cluster Terzan 5},
  journal = {Astronomy \& Astrophysics},
  volume  = {686},
  pages   = {A166},
  year    = {2024},
  doi     = {10.1051/0004-6361/202449303}
}

@article{wpq+24,
  author  = {{Wu}, Yuxiao and {Pan}, Zhichen and {Qian}, Lei and others},
  title   = {The Discovery of Three Pulsars in the Globular Cluster M15 with FAST},
  journal = {The Astrophysical Journal Letters},
  volume  = {974},
  pages   = {L23},
  year    = {2024},
  doi     = {10.3847/2041-8213/ad7b9e}
}

@article{dcm+23,
  author  = {{Dong}, Fengqiu Adam and {Crowter}, Kathryn and {Meyers}, Bradley W. and others},
  title   = {The second set of pulsar discoveries by CHIME/FRB/Pulsar: 14 Rotating Radio Transients and 7 pulsars},
  year    = {2023},
  eprint  = {2210.09172},
  archivePrefix = {arXiv},
  primaryClass = {astro-ph.HE}
}

@article{shw+23,
  author  = {{Su}, W. Q. and {Han}, J. L. and {Wang}, P. F. and others},
  title   = {The FAST Galactic Plane Pulsar Snapshot Survey - III. Timing results of 30 newly discovered pulsars},
  journal = {Monthly Notices of the Royal Astronomical Society},
  volume  = {526},
  pages   = {2645--2656},
  year    = {2023},
  doi     = {10.1093/mnras/stad2159}
}

@article{wyw+23,
  author  = {Wu, Q. D. and Yuan, J. P. and Wang, N. and others},
  title   = {Follow-up timing of 24 pulsars discovered in commensal radio astronomy FAST survey},
  journal = {Monthly Notices of the Royal Astronomical Society},
  volume  = {522},
  pages   = {5152--5164},
  year    = {2023},
  doi     = {10.1093/mnras/stad1323}
}

@article{psf+22,
  author  = {{Parent}, E. and {Sewalls}, H. and {Freire}, P. C. C. and others},
  title   = {Study of 72 Pulsars Discovered in the PALFA Survey: Timing Analysis, Glitch Activity, Emission Variability, and a Pulsar in an Eccentric Binary},
  journal = {The Astrophysical Journal},
  volume  = {924},
  pages   = {135},
  year    = {2022},
  doi     = {10.3847/1538-4357/ac375d}
}

@article{Deutsch1955,
  author  = {Deutsch, A. J.},
  title   = {The electromagnetic field of an idealized star in rigid rotation in vacuo},
  journal = {Annales d'Astrophysique},
  volume  = {18},
  pages   = {1--10},
  year    = {1955}
}

@article{Muslimov1986SvA,
  author  = {Muslimov, A. G. and Tsygan, A. I.},
  title   = {Electric fields of neutron stars with an arbitrary inclination of magnetic axes},
  journal = {Soviet Astronomy},
  volume  = {30},
  pages   = {567},
  year    = {1986}
}

@article{Konno00,
  author  = {Konno, K. and Kojima, Y.},
  title   = {General relativistic effects of gravity in quantum mechanics: a case of ultra-relativistic particles},
  journal = {Progress of Theoretical Physics},
  volume  = {104},
  pages   = {1117},
  year    = {2000}
}

@article{Hakimov13,
  author  = {Hakimov, A. and Abdujabbarov, A. and Ahmedov, B.},
  title   = {Magnetic fields of spherical compact stars in modified theories of gravity: $f(R)$ type gravity and Ho{\v{r}}ava-Lifshitz gravity},
  journal = {Physical Review D},
  volume  = {88},
  pages   = {024008},
  year    = {2013},
  doi     = {10.1103/PhysRevD.88.024008}
}

@article{RezzollaAhmedov2004,
  author  = {Rezzolla, L. and Ahmedov, B. J.},
  title   = {Electromagnetic fields of a slowly rotating magnetized star in general relativity},
  journal = {Monthly Notices of the Royal Astronomical Society},
  volume  = {352},
  pages   = {1161},
  year    = {2004},
  note    = {Related to the 2001 series of papers}
}

@article{Zhang2004,
  author  = {Zhang, L. and Cheng, K. S. and Jiang, Z. J. and Leung, P.},
  title   = {Gamma-Ray Luminosity and Death Lines of Pulsars with Outer Gaps},
  journal = {The Astrophysical Journal},
  volume  = {604},
  pages   = {317},
  year    = {2004},
  doi     = {10.1086/381794}
}

@article{Hewish68,
	Adsurl = {http://adsabs.harvard.edu/abs/1968Natur.217..709H},
	Author = {{Hewish}, A. and {Bell}, S.~J. and {Pilkington}, J.~D.~H. and {Scott}, P.~F. and {Collins}, R.~A.},
	Doi = {10.1038/217709a0},
	Journal = {Nature},
	Month = feb,
	Pages = {709-713},
	Title = {{Observation of a Rapidly Pulsating Radio Source}},
	Volume = 217,
	Year = 1968}

@ARTICLE{Beloborodov08,
   author = {{Beloborodov}, A.~M.},
    title = "{Polar-Cap Accelerator and Radio Emission from Pulsars}",
  journal = {Astrophys. J.},
archivePrefix = "arXiv",
   eprint = {0710.0920},
     year = 2008,
    month = aug,
   volume = 683,
      eid = {L41},
    pages = {L41},
      doi = {10.1086/590079},
   adsurl = {http://adsabs.harvard.edu/abs/2008ApJ...683L..41B}
}

@article{Morozova08,
	Adsurl = {http://adsabs.harvard.edu/abs/2008ApJ...684.1359M},
	Archiveprefix = {arXiv},
	Author = {{Morozova}, V.~S. and {Ahmedov}, B.~J. and {Kagramanova}, V.~G.},
	Doi = {10.1086/590322},
	Eprint = {0806.2376},
	Journal = {Astrophys. J.},
	Month = sep,
	Pages = {1359-1365},
	Title = {{General Relativistic Effects of Gravitomagnetic Charge on Pulsar Magnetospheres and Particle Acceleration in the Polar Cap}},
	Volume = 684,
	Year = 2008}

\end{document}